\documentclass[amsmath,nofootinbib,amssymb,aps,twocolumn]{revtex4}  % PRD

\usepackage{amsmath}
\usepackage{amssymb}
\usepackage{amsthm}
\usepackage{psfrag}
\usepackage{graphicx}
\usepackage{hyperref}
\usepackage{color}

\newcommand{\be}{\begin{equation}}
\newcommand{\ee}{\end{equation}}

\newcommand{\pd}{\partial}
\newcommand\m{\mu}

\newcommand\n{\nu}
\renewcommand\r{\rho}
\newcommand\s{\sigma}
\renewcommand\a{\alpha}
\renewcommand\b{\beta}
\renewcommand\l{\lambda}
\newcommand{\T}{\Theta}
\newcommand{\HH}{{\cal H}}
\newcommand{\Gc}{G_{\text{cos}}}

\def\d{\partial}
\newcommand{\bseq}{\begin{subequations}}
\newcommand{\eseq}{\end{subequations}}

\newcommand{\di}{\mathrm d}
\renewcommand{\H}{\mathcal H}

\begin{document}

\begin{flushright}  % PRD
CERN-PH-TH/2013-085, LAPTH-022/13
\end{flushright}
\vskip -0.9cm
\title{Cosmological constraints on Lorentz violating dark energy}
\author{B. Audren\,$^{a}$, D. Blas\,$^{b},$ J. Lesgourgues\,$^{a,b,c}$,
  S. Sibiryakov\,$^{d,e}$}
  \affiliation{{$^a$} \it FSB/ITP/LPPC,
 \'Ecole Polytechnique F\'ed\'erale de Lausanne,\\
 \normalsize\it CH-1015, Lausanne, Switzerland}
  \affiliation{{$^b$} \it Theory Group, Physics Department, CERN,
   CH-1211 Geneva 23, Switzerland}
 \affiliation{{$^c$} \it LAPTh (CNRS -Université de Savoie), BP 110,
  F-74941 Annecy-le-Vieux Cedex, France}
    \affiliation{{$^d$} \it  Institute for Nuclear Research of the
Russian Academy of Sciences,
  60th October Anniversary Prospect, 7a, 117312 Moscow, Russia}
\affiliation{{$^e$} \it  Faculty of Physics, Moscow State University, Vorobjevy Gory,
 119991 Moscow, Russia}

\begin{abstract} % PRD
The role of Lorentz invariance as a fundamental symmetry of nature has
been lately reconsidered in different approaches to quantum gravity.
It is thus natural to study whether other puzzles of physics may be
solved within these proposals. This may be the case for the
cosmological constant problem. Indeed, it has been shown that breaking
Lorentz invariance provides Lagrangians that can drive the current
acceleration of the universe without experiencing large corrections
from ultraviolet physics. In this work, we focus on the simplest
model of this type, called $\T$CDM, and study its cosmological
implications in detail. At the background level, this model cannot be
distinguished from $\Lambda$CDM. The differences appear at the level
of perturbations. We show that in $\T$CDM, the spectrum of CMB
anisotropies and matter fluctuations may be affected by a rescaling of
the gravitational constant in the Poisson equation, by the presence of
extra contributions to the anisotropic stress, and finally by the
existence of extra clustering degrees of freedom. To explore these
modifications accurately, we modify the Boltzmann code  {\sc class}.
We then use the parameter inference code {\sc Monte Python} to
confront $\T$CDM with data from WMAP-7, SPT and WiggleZ. We obtain
strong bounds on the parameters accounting for deviations from
$\Lambda$CDM.  In particular, we find that the discrepancy between the
gravitational constants appearing in the Poisson and Friedmann
equations is constrained at the level 1.8\%.
\end{abstract}

\maketitle

%\newpage

%%%%%%%%%%%%%%%%%%%%%%%%%
\section{Introduction}
%%%%%%%%%%%%%%%%%%%%%%%%%

Explaining the origin of the current acceleration of the universe is
one of the biggest challenges in cosmology. Despite the great
successes of the $\Lambda$CDM paradigm,  it is hard to accept that a
very fine-tuned form of otherwise invisible energy governs nowadays
the behavior of the universe at the largest scales. A whole plethora
of models are proposed as alternatives to this situation. These are
known as \emph{quintessence} or \emph{dark energy} models
\cite{DEBook,Amendola:2012ys,Copeland:2006wr}. The motivations and
domain of applicability of various proposals are quite diverse:
whereas some of them are intended as mere phenomenological models only
valid for cosmology, others are rooted in theoretical considerations
and are testable by different types of experiments. From the
theoretical viewpoint the preferred models are those addressing (at
least partially) the naturalness problem of the cosmological constant
without introducing any additional fine-tunings.  In this paper we
will study in detail the  $\T$CDM proposal of Ref.~\cite{Blas:2011en}
that has such a property. This model is well-motivated theoretically
and has a rich phenomenology that may distinguish it from
$\Lambda$CDM. 

The $\T$CDM model is based on the idea that the Lorentz invariance
observed in the Standard Model of particle physics is an emergent
phenomenon and does not correspond to a symmetry of nature. This
concept is motivated by attempts to find complete theories of quantum
gravity, but it can also be considered independently. There exist two
different approaches: on one hand Einstein-aether theory
\cite{Jacobson:2000xp,Jacobson:2008aj} provides a phenomenological
description of gravity with broken Lorentz symmetry at large
distances; on the other hand Ho\v rava gravity
\cite{Horava:2009uw,Blas:2009qj} invokes a violation of Lorentz
invariance at any scale to improve the quantum properties of
gravitational theories.  The restriction of Ho\v rava gravity to
operators  with the lowest dimension can be considered as an effective
theory by itself, called the \emph{khronometric} theory
\cite{Blas:2010hb}.  The two proposals are very similar at large
distances: the khronometric theory can be viewed as a constrained
version of the Einstein-aether theory, and in many situations the
predictions of the two theories coincide
\cite{Jacobson:2010mx,Blas:2010hb}.  The analysis presented in this
work is applicable to all these classes of models.  However, in the
last step consisting in deriving the actual observational bounds on
the model parameters, we will restrict to the khronometric case. 
 
Several implications of the above models for cosmology have been
studied in the past,  see e.g.
\cite{Zuntz:2008zz,Kobayashi:2010eh,ArmendarizPicon:2010rs,
Carroll:2004ai,Nakashima:2011fu,Li:2007vz,Zlosnik:2007bu}.  For the
evolution of the Universe as a whole, the consequences of the minimal
models\footnote{By this term we refer to the models with a minimal
  number of additional degrees of freedom in the gravity sector, and
  the simplest kinetic action (Einstein-aether and khronometric
  theories) \cite{Jacobson:2008aj,Blas:2010hb}.} 
are rather trivial. They are compatible with
Friedmann--Robertson--Walker (FRW) solutions that differ from the
standard case only by a renormalization of the gravitational constant
away from the value measured in local tests of Newton's law.  In other
words, those models do not modify the form of the Friedmann equation.
The background evolution becomes more interesting  when non-minimal
models are considered. In this case, the breaking of Lorentz
invariance allows for Lagrangians which can change the expansion
history of the universe and provide new alternatives for inflationary
dynamics, dark matter and dark energy
\cite{Donnelly:2010cr,Blas:2011en,Zlosnik:2007bu,Zuntz:2008zz}. Some
criteria are necessary to identify the most interesting cases.  In
this work we will be concerned with the issue of dark energy, for
which one would like to find a simple model described by a Lagrangian
with high cutoff scale, that provides a mechanism to accelerate the
universe insensitive to UV  corrections\footnote{We leave aside the
  ``old cosmological constant problem'': to find a mechanism imposing
  a null vacuum energy (see however the related comments in
  \cite{Blas:2011en}.)} 
and distinguishable from $\Lambda$CDM. The $\T$CDM model of
Ref.~\cite{Blas:2011en} meets these requirements.

Some consequences of $\T$CDM have been already discussed in
Ref.~\cite{Blas:2011en}.  In the present work we describe the
observable physical effects of the model on cosmic microwave
background (CMB)  anisotropies and on the matter power spectrum (at
the linear level).  We provide a detailed discussion of such effects,
that we computed accurately with a modified version of the flexible
Boltzmann code {\sc class} \cite{Blas:2011rf}. We then compare the
$\T$CDM model to recent CMB and Large Scale Structure (LSS) data using
the Monte Carlo parameter inference code {\sc Monte Python}
\cite{Audren:2012wb}.
 
Our work is organized as follows: in Sec.~\ref{sec:Lagr} we describe
the $\T$CDM model and discuss the constraints not related to
cosmology. In Sec.~\ref{sec:cosmo} we study the background evolution
and derive linear equations for perturbations around the FRW
background. The results presented in this section are complementary
to those in Ref.~\cite{Blas:2011en}. They hold in a different gauge,
and refer to different conventions and parametrizations, found to be
more suitable for the numerical implementation.  The qualitative
effects of $\T$CDM on the CMB and matter power spectrum are described
in Sec.~\ref{sec:obser}. We present constraints from CMB and LSS data
in Sec.~\ref{sec:data}, and expose our conclusions in
Sec.~\ref{sec:concl}. Appendix~\ref{sec:app} contains the derivation
of initial conditions for cosmological perturbations in $\T$CDM.

%%%%%%%%%%%%%%%%%%%%%%%%%
\section{Lorentz breaking theories of gravity and $\T$CDM}\label{sec:Lagr}
%%%%%%%%%%%%%%%%%%%%%%%%%

Lorentz invariance is one of the best tested symmetries of the
Standard model of particle physics \cite{Kostelecky:2008ts}.  It is
also a fundamental ingredient of the theory of general relativity,
that provides a very successful description of gravitational
interactions over a huge range of scales. However, it is not known
how to directly promote general relativity to a complete quantum
theory, which points towards the necessity to consider alternatives.
Independently of this, modifications to general relativity are
currently being considered in the area of cosmology. The rationale
behind these modifications is the possibility to use the wealth of
cosmological data to learn how gravitation behaves at the largest
accessible distances and, hopefully, shed some light on the mechanism
responsible for the accelerated expansion of the universe.

These two lines of research converge if one assumes that Lorentz
invariance is not a symmetry of the gravitational sector. This idea
opens the possibility to construct gravitational theories with better
quantum behavior than general relativity \cite{Horava:2009uw}.  This
is achieved at the price of introducing new degrees of freedom that
modify the laws of gravity at all distances, including those relevant
for cosmology \cite{Blas:2009qj,Blas:2010hb}. Even before this
top-down approach had been initiated, the bottom-up Einstein-aether
model \cite{Jacobson:2000xp} was proposed as a way to capture the
large-distance effects of a putative Lorentz violation due to quantum
gravity.  In this work, we will consider theories where Lorentz
violation is described by a preferred time-like vector field $u_\m$
defined at every point of space-time, which includes Einstein-aether
theory and Ho\v rava gravity. For the latter,  we will focus only on
its low-energy form, the khronometric theory \cite{Blas:2010hb}.  The
vector field will be normalized to\footnote{We use the $(-+++)$
  signature for the metric. This differs from most of the previous
  works in the field, but is common in cosmology.}
\be
\label{constr}
u_\m u^\m=-1.
\ee
The presence of this dynamical field allows to use for the description
of Lorentz violation the same language as for spontaneous symmetry
breaking.  Gravitational physics at large distances is governed by the
following covariant action for $g_{\m\n}$ and $u_\m$, featuring a
minimal number of derivatives:
\be
\begin{split}
\label{aetheract}
S_{[ \mathrm{EH}u]}=\frac{M_0^2}{2}\int \di^4x \sqrt{-g}
\large[R\,-\,&K^{\m\n}_{\phantom{\m\n}\s\r}\nabla_\m u^\s\nabla_\nu u^\rho\\
&\quad\quad+l (u_\m u^\m+1)\large]\;,
\end{split}
\ee
where
\be
\label{Kmnsr}
K^{\m\n}_{~~~\s\r}\equiv c_1g^{\m\n}g_{\s\r}+c_2\delta_\s^\m\delta_\r^\n
+c_3\delta_\r^\m\delta_\s^\n-c_4u^\m u^\n g_{\s\r},
\ee
and $l$ is a Lagrange multiplier that enforces the unit-norm
constraint. Equation~(\ref{aetheract}) includes the Einstein-Hilbert
(EH) term evaluated with the metric $g_{\m\n}$. The parameter $M_0$ is
proportional to the Planck mass (cf. \eqref{gn}) while the
dimensionless constants $c_a$, $a=1,2,3,4$, characterize the strength
of the interaction of the aether  $u_\m$ with gravity. This is the
action of the Einstein-aether model
\cite{Jacobson:2000xp,Jacobson:2008aj}.

The khronometric case corresponds to the situation where $u_\m$ is
defined as a vector normal to a foliation consisting of the level
surfaces of the {\it khronon} field $\varphi$, 
\be
\label{ukh}
u_\m\equiv\frac{\pd_\m \varphi}{\sqrt{-\nabla^\n \varphi \pd_\n \varphi}}.
\ee
In this case, the constraint (\ref{constr}) is satisfied identically
and the first term of (\ref{Kmnsr}) can be expressed as a linear
combination of the last two terms. Thus, $K^{\m\n}_{~~~\s\r}$
reduces to its last three terms with coefficients 
\be
\label{khpar}
\l\equiv c_2,~~~\beta\equiv c_3+c_1,~~~\alpha\equiv c_4+c_1\;.
\ee
At the level of linear perturbations, the khronometric theory differs
from the Einstein-aether theory by the number of propagating degrees
of freedom apart from the spin-two mode of the graviton. While the
aether in general describes vector and scalar excitations, the
khronometric case only contains scalars. However, the scalar sectors
of the two theories are equivalent and are completely characterized by
the three parameters (\ref{khpar}). 
 
Once the couplings to matter are specified, the constants $c_a$ are
constrained by various considerations ranging from theoretical
requirements to observational tests. First, to very good precision,
the Standard Model fields must couple only to $g_{\m\n}$ and not to
$u_\m$, as required by Lorentz invariance in this sector\footnote{More
  generally, a universal coupling of Standard Model fields to a fixed
  combination of $g_{\m\n}$ and $u_\m$ is allowed. This reduces to the
  previous case by a redefinition of the metric \cite{Foster:2005ec}.} 
\cite{Kostelecky:2008ts}.  This decoupling presents a serious
challenge to the proposal, but it is conceivable to achieve it either
by imposing extra symmetries, e.g. supersymmetry
\cite{GrootNibbelink:2004za,Pujolas:2011sk}, or through a
renormalization group running \cite{PS2}.  Next, there are
restrictions imposed by the stability of Minkowski spacetime
\cite{Jacobson:2008aj}. In particular, the requirement that the scalar
mode is neither a ghost nor a tachyon field amounts to the constraints
\be
\label{stab}
0<\a<2~,~~~\b+\l>0\;.
\ee
Stringent bounds come from observations of the Solar System dynamics,
which can be used to place constraints on post-Newtonian (PPN)
parameters. Two of these parameters, denoted by $\a_1^{PPN}$ and
$\a_2^{PPN}$, describe the effects of Lorentz violation. In general,
they are different from zero (their value in general relativity) both
in the Einstein-aether and khronometric theory; we refer the reader to
\cite{Jacobson:2008aj,Blas:2011zd} for explicit relations with the
constants $c_a$. Assuming no cancellations in these formulae, one
obtains experimental bounds of the order of
\be
\label{eq:bounda}
|c_{a}|\lesssim 10^{-7}\;.
\ee
However, there are regions in the space of model parameters $(\alpha$,
$\beta$, $\lambda$) in which $\a_{1,2}^{PPN}$ vanish and Solar System
tests are automatically satisfied, with Newton's constant given by 
\be
\label{gn}
G_{N}\equiv\frac{1}{8\pi M_0^2(1-\alpha/2)}\;.
\ee
This requires\footnote{The difference between the models is due to the
  contributions to $\a_{1,2}^{PPN}$ from the vector polarizations
  that are present only in the Einstein-aether case.}
\bseq
\label{abl}
\begin{align}
\label{abkhronon}
&\a=2\b~~ \text{for the khronometric model,}
\\
\label{ablaether}
&\alpha=-(3\lambda+\beta)~~ \text{for the Einstein-aether.}
\end{align} 
\eseq
In these cases, much weaker bounds can be inferred from gravitational
wave emission in binary systems \cite{Foster:2006az,Blas:2011zd},
giving
\be
\label{eq:boundb}
|c_{a}|\lesssim 10^{-2}\;,
\ee
and from the form of black hole solutions, imposing inequalities
detailed in~\cite{Barausse:2011pu}.  We will see that in the case of
the khronometric model, the bounds that can be inferred from cosmology
are competitive with (\ref{eq:boundb}), but not with
(\ref{eq:bounda}). Thus we will impose the relation (\ref{abkhronon})
when searching for the allowed parameter space.  Previous studies of
the cosmological effects of Lorentz violation
\cite{Zuntz:2008zz,Li:2007vz} focused on the Einstein-aether model.
In those works, the relation (\ref{ablaether}) was enforced to avoid
the PPN constraints. As we will explain, the relation
(\ref{ablaether}) also incidentally suppresses the leading effects in
cosmology, unlike the relation (\ref{abkhronon}). This explains why
the bounds obtained in those studies are rather mild. 

The implications of the action (\ref{aetheract}) for the background
evolution of an homogeneous and isotropic universe are minimal. If all
matter components are universally coupled\footnote{A priori, there is
  no reason why this must be true for the dark matter. Still, as shown
  in \cite{Blas:2012vn}, even allowing for non-universal interactions
  between the dark matter and the aether does not change the
  background evolution. As our main focus in this paper is dark
  energy, we will assume that the dark matter has standard properties
  (namely, that it is a pressureless fluid, described at the
  fundamental level by a Lorentz invariant Lagrangian, and with
  universal coupling to $g_{\m\n}$).}
to $g_{\m\n}$, the only difference  with respect to general relativity
is that the Friedmann equation involves a renormalized gravitational
constant
\be
\label{gc}
G_{cos}\equiv\frac{1}{8\pi M_0^2(1+\beta/2+3\lambda/2)}\;
\ee
 differing from the value $G_N$ measured e.g. on earth or in the Solar
 System\footnote{The analysis of Big Bang Nucleosynthesis
   \cite{Carroll:2004ai} sets a bound on the relative difference
   between the two, $|G_{cos}/G_N-1|\leq 0.13$.},
given by Eq.~(\ref{gn}).  In order to modify the expansion history of
the universe, and find a candidate for dark energy, we need to add a
new ingredient to the model. We want to do it in a way that preserves
the simplicity of the proposal and its validity as a low-energy
effective field theory.  This is achieved by supplementing the action
(\ref{aetheract}) with a new field $\T$ invariant under the shift
symmetry,
\be
\label{eq:shift}
\T\mapsto \T+const.
\ee
The low-energy action for this field is\footnote{ A similar action
  with an additional potential term for $\T$ breaking the shift
  symmetry (\ref{eq:shift}) was considered in \cite{Donnelly:2010cr}.}
\be\begin{split}
\label{Thetaact}
S_{[\Theta]}=\int \di^4x \sqrt{-g}& \bigg(-\frac{g^{\m\n}\d_\m\Theta\d_\n\Theta}{2}\\
~~~~~&+
\kappa\frac{(u^\mu\d_\mu\Theta)^2}{2}
-\mu^2u^\mu\d_\mu\Theta\bigg)\;,
\end{split}
\ee
and involves two free parameters ($\kappa$, $\mu$).  We will refer to
the model resulting from the combination of the actions
(\ref{Thetaact}) and (\ref{aetheract}) (with a universal coupling
between matter fields and $g_{\m\n}$) as $\T$CDM \cite{Blas:2011en}.
The cosmological constant term is set to be exactly zero\footnote{One
  can entertain the possibility of finding a mechanism that would
  enforce the cancellation of the vacuum energy induced by quantum
  loops \cite{Blas:2011en}. However, at present, we are not aware of
  any such mechanism. Thus, the vanishing of the cosmological constant
  should be taken merely as an assumption.}. 
Two comments are in order. First, as an effective field theory,
$\T$CDM provides a valid description of physics up to  a cutoff scale
of the order of $\Lambda_c\sim \sqrt{c_a}M_P$.  The bounds
(\ref{eq:bounda}) or (\ref{eq:boundb}) show that this cutoff scale may
be only a few orders of magnitude below the Planck mass.  Furthermore,
in the khronometric case, the theory has a potential UV completion,
since it is a sub-case of Ho\v rava gravity \cite{Horava:2009uw}.
Second, in the whole $\T$CDM action, only the last operator in
(\ref{Thetaact}) breaks the discrete symmetry $\T\to -\T$. This
implies that from the viewpoint of the effective field theory, it is
self-consistent to choose the coefficient in front of this operator to
be much smaller than the UV cutoff. In spite of $\m$ being
dimensionfull, the above symmetry guarantees that it is renormalized
multiplicatively, and that no dangerous contributions proportional to
$\Lambda_c$ appear. In other words, the smallness of $\m$ is
technically natural.  This last observation is very important, since
we are going to see that in the $\T$CDM  model, $\mu$ sets the scale
of the current cosmic acceleration.

%%%%%%%%%%%%%%%%%%%%%%%%%
\section{Cosmological solutions of $\T$CDM}\label{sec:cosmo}
%%%%%%%%%%%%%%%%%%%%%%%%%

We are interested in describing the evolution of perturbations around
homogeneous and isotropic solutions in the $\T$CDM model.  We focus on
scalar perturbations, and refer the reader to
\cite{Nakashima:2011fu,ArmendarizPicon:2010rs,Lim:2004js} for possible
effects of vector and tensor modes. In the synchronous gauge, the
perturbed FLRW metric has the form, 
\be
\label{FRW}
\begin{split}
&g_{00}=-a(\tau)^2, \quad g_{0i}=0, \quad\\
&g_{ij}=a(\tau)^2\left[\delta_{ij}+\frac{\pd_i\pd_j}{\Delta} h+6\left(\frac{\pd_i\pd_j}{\Delta}-\frac{1}{3}\delta_{ij}\right)\eta\right].
\end{split}
\ee
Besides, for $\Theta$ and the khronon (or the longitudinal component
of $u_\m$ in the more general Einstein-aether case) we introduce 
\be
\Theta=\bar \Theta(\tau)+\xi, \quad \varphi=\tau+\chi.
\ee
The matter components are assumed to be cold dark matter ($cdm$),
photons ($\gamma$), neutrinos ($\n$) and baryons ($b$).  We describe
these components at the same level of approximation as in
Ref.~\cite{Ma:1995ey}.  In particular, the dark matter is treated as a
pressureless fluid universally coupled to the metric $g_{\m\n}$.  We
assume that it is comoving with the gauge, in order to eliminate the
well-known residual freedom in the synchronous gauge.  We now discuss
how the gravitational equations are modified in $\T$CDM.  These
equations were derived in \cite{Blas:2011en} using the conformal
Newtonian gauge.  Here we will rewrite the linearized equations in the
synchronous gauge \eqref{FRW}, and in a form optimized for numerical
study with the Boltzmann code {\sc class} \cite{Blas:2011rf}. 

%%%%%%%%%%%%%%%%%%%%%%%%%%%
\subsection{Background evolution}
%%%%%%%%%%%%%%%%%%%%%%%%%%%

Deriving the equation of motion for $\T$ from (\ref{Thetaact}), one
finds that the homogeneous part $\bar\T(t)$ evolves as
\be
\label{eq:backg}
\dot{\bar \Theta}=-\frac{\m^2  a}{1+\kappa}+\frac{C}{a^2}\;,
\ee
where $C$ is an integration constant. Substituting this solution into
the Friedmann equation (derived from the combination of the actions
(\ref{aetheract}) and (\ref{Thetaact})) yields
\be
\label{eq:Friedm}
H^2
=\frac{8\pi G_{cos}}{3}\left(\rho_\m+\rho_s+\rho_d+\sum_{\rm other}\rho_n\right),
\ee
where $H\equiv \dot a/a^2$ and $G_{cos}$ is given by (\ref{gc}). The
first three contributions in the brackets come from the
energy-momentum tensor of the $\T$-field and have the form, 
\be
\rho_\m\equiv\frac{\m^4}{2(1+\kappa)}\;, 
\quad \rho_s\equiv\frac{C^2(1+\kappa)}{2a^6}\;,\quad
\rho_d=-\frac{\m^2C}{a^3}\;,
\ee
while $\r_n$, $n=cdm,\gamma,\n,b$, stand for the densities of the
standard matter components of the Universe.

Let us analyze Eq.~(\ref{eq:Friedm}). Recall that there is no
cosmological constant at the fundamental level in the model. Instead,
the first term in (\ref{eq:Friedm}) plays the same role with an energy
scale set\footnote{Throughout the paper we assume that the combination
  $1+\kappa$ is of order one.}
by $\mu$.  As emphasized above, $\m$ does not receive large radiative
corrections, and thus this source of dark energy can naturally have an
energy scale completely unrelated to the cutoff of the theory.  The
second term has the form of the contribution of stiff matter.  Not to
spoil Big Bang Nucleosynthesis (BBN), $\rho_s$ must be smaller than
about 30 times the density of radiation  in the Universe at
temperatures of the order of 10 MeV \cite{Dutta:2010cu}.  Thus,
assuming that this contribution was already present at BBN\footnote{We 
  do not specify the origin of the field $\T$ in this work. One
  possibility is to identify $\T$ with the Goldstone boson of a
  spontaneously broken global symmetry.  Then, the above assumption
  amounts to stating that the corresponding phase transition occurs
  before BBN. If it happens later, the picture may change, but it is
  hard to see how this option can be incorporated in a viable
  cosmological scenario.},
due to its rapid decrease, it is completely negligible at later
epochs.  The third term in (\ref{eq:Friedm}) behaves as the energy
density of dust\footnote{Note though, that its sign can be negative
  depending on the sign of $C$.},
so one might be tempted to identify it with the dark matter. However,
being the geometric mean of the first two, this term is always
subdominant and cannot contribute a significant fraction of dark
matter. Given these considerations, we will set $C=0$ henceforth.

%%%%%%%%%%%%%%%%%%%%%%%%%%%
 \subsection{Cosmological perturbations\label{ssec:pk}}
%%%%%%%%%%%%%%%%%%%%%%%%%%%%
\label{ssec:cosper}

Let us introduce two time scales that appear in the analysis of the
linear perturbations,
\be
\label{scales}
\tau^{-1}_\a\equiv\sqrt\frac{8\pi G_{cos}}{\a}\,\dot{\bar\T}\;,\quad
H_\a\equiv\frac{H_0}{\sqrt{\a}}\;,
\ee
where $H_0 = 100\ \mathrm{h \ km\ s^{-1}\ Mpc^{-1}}$ is the current
value of the Hubble constant today.  It is also convenient to rescale
the $\T$-fluctuation defining
\be
\label{tildee}
\tilde \xi  \equiv \frac{\sqrt{8 \pi \Gc}}{H_0} \xi\;,
\ee
so that both fields $(\chi,\tilde  \xi)$ have the dimension of time.
The equations of motion for $\chi$ and $\tilde \xi$ read
\bseq
\label{eqchixi}
\begin{align}
  \ddot{\chi} =& - 2 {\cal H} \dot{\chi} \nonumber\\
  -\bigg[&k^2c_\chi^2 +  (1+B){\cal H}^2 +(1-B)\dot{\cal H} +
  \frac{G_0}{\Gc \tau^{2}_\a}    \bigg] \chi \nonumber \\
  &+\frac{G_0}{\Gc}\frac{H_\alpha}{\tau_\a} \tilde\xi 
-\frac{c_\chi^2}{2}\dot h - 2\frac{\beta}{\alpha}\dot \eta,
  \label{eq:chi}\\
\label{eq:xi}
  \ddot{\tilde \xi} =& -2\mathcal{H} \dot{ \tilde \xi} -
  k^2c_\Theta^2\tilde{\xi}+\frac{k^2c_\Theta^2}{H_\a\tau_\a}\chi, 
\end{align}
\eseq
where we introduced the notations 
\be
\begin{split}
& G_0 \equiv \frac{1}{8 \pi M_0^2}\;, \quad
\frac{G_{cos}}{G_0} \equiv\frac{1}{1 + \beta/2 + 3 \lambda/2}\;, \\
&B\equiv\frac{\beta+3\lambda}{\alpha}\;,\quad
c_\chi^2\equiv \frac{\beta+\lambda}{\alpha}\;,
\quad c^2_\T\equiv \frac{1}{1+\kappa}\;,
\end{split}
\ee
and
\be
\HH\equiv a H.
\ee
Note that the constants $c_\chi$, $c_\T$ represent the sound speeds of the
fields $\chi$ and $\xi$ at relatively short wavelengths, where these modes
decouple from each other \cite{Blas:2011en}.  These two velocities are constant
in time and we will treat them as quantities of order one, which is their
natural order of magnitude from the point of view of effective field theory.
Note that in general they can exceed one, so that the fields can be
superluminal. In Lorentz violating theories, this does not lead to any causal
paradoxes, see e.g. the discussion in Ref.~\cite{Mattingly:2005re}.

We now present the linearized Einstein equations. There are four
equations in the scalar sector corresponding to different components
of the Einstein tensor 
$G^\m_{\phantom{\m}\n}\equiv R^\m_{\phantom{\m}\n}-\frac{1}{2}
\delta^\m_{\phantom{\m}\n}R$.
Only two of these equations are independent, but we write all of them
for completeness. For the $\delta G^0_{\phantom{0}0}$ component we
find
\begin{align}
  \bigg(k^2\eta - \frac{1}{2}\frac{G_0}{\Gc}\mathcal H \dot h\bigg) 
= -4\pi a^2 {G_0}\sum_n \rho_n\delta_n,
  \label{eq:T00}
\end{align}
with
\be
\begin{split}
 \sum_i \rho_i\delta_i \equiv& \sum_{\rm other} \rho_i\delta_i +
 \frac{\a}{8\pi a^2 G_{cos}c_\Theta^2}\frac{H_\a}{\tau_\a}\dot{\tilde
 \xi}\\
  &+ \frac{\alpha\, k^2}{8\pi a^2G_0}  (\mathcal H (1-B)\chi + \dot\chi ).
  \label{}
\end{split}
\ee
Here and in what follows, the label `other' refers to contributions
from the standard matter components, whose form can be found in
\cite{Ma:1995ey}. For the $\pd_i\delta G^0_{\phantom{0}i}$ part we
find 
\begin{align}
  2k^2(1-\beta)\dot\eta - \frac{\a\,  c_\chi^2}{2}k^2\dot h = 8 \pi a^2{G_0} \sum_n(\rho_n +p_n)\theta_n, 
  \label{eq:etadot}
\end{align}
with
\be
\label{eq:theta}
\begin{split}
\sum_n(\rho_n + p_n)\theta_n \equiv\sum_{\rm other}(\rho_n +  p_n)\theta_n +  \frac{\a\,  c_\chi^2}{8\pi a^2G_0}k^4\chi.
\end{split}
\ee
The $\delta G^i_{\phantom{i}i}$ equation reads
\begin{align}
\label{eq:Tii}
  \ddot h &= -2\mathcal H \dot h + 2\frac{\Gc}{G_0} k^2 \eta-24\pi G_{cos}a^2\sum_i\delta p_n,
\end{align}
where
\be
\label{eq:press}
\sum_n\delta p_n\equiv \sum_{\rm other}\delta p_n +  \frac{\a\,
  B}{24\pi a^2G_0}k^2(\dot \chi + 2\mathcal H\chi)\;. 
\ee
Finally, the  $\pd_i\pd_j\delta G^i_{j}$ equation yields
\begin{align}
\label{eq:tracefree}
  (1-\beta)\big(\ddot h + 6\ddot \eta + 2\mathcal H (\dot h +
  6\dot\eta)\big) 
&- 2k^2\eta =\nonumber\\-24\pi a^2{G_0}
  &\sum_n( \rho_n +  p_n)\sigma_n,
\end{align}
with
\be
\sum_n(\rho_n + p_n)\sigma_n \equiv\sum_{\mathrm{other}}( \rho_n + p_n)\sigma_n  - \frac{\beta k^2}{12\pi a^2G_0}(\dot\chi + 2\mathcal H \chi).
\ee
In the above expressions the pressure fluctuations $\delta p_n$, the
velocity divergences $\theta_n$ and the shear potentials $\sigma_n$
are defined in the standard way, see \cite{Ma:1995ey}; the equations
for perturbations in the matter components can be found in the same
reference. The linearized equations must be supplemented  by suitable
initial conditions. The latter are derived in Appendix~\ref{sec:app}.   

For simplicity, we are going to set $c_\T=1$ in the numerical
simulations. This prescription is equivalent to fixing $\kappa=0$, and
leaves us with four free fundamental parameters ($\alpha$, $\beta$,
$\lambda$, $\mu$).  A different choice for $c_\T$ would not affect
qualitatively the evolution of perturbations, unless $c_\T$ is very
small. In that case, the field $\tilde{\xi}$ could in principle
cluster on scales much smaller than the Hubble radius, but we will not
consider this situation in the present work\footnote{A small value of
  $c_\Theta$ corresponds to the near cancellation of the kinetic terms
  for $\tilde{\xi}$ coming from the first and second terms in
  Eq.~(\ref{Thetaact}) - a situation that appears fine-tuned from the
  effective field theory perspective.}.

We can compare the number of free parameters in this model with that
in an ordinary $\Lambda$CDM model sharing the same background
evolution. The parameters ($\Omega_\Lambda$, $H_0$) of $\Lambda$CDM
can be mapped onto the parameters ($\mu$, $H_0$) of the $\T$CDM model.
Hence the latter features only three additional parameters ($\alpha$,
$\beta$, $\lambda$), reducing to only two independent parameters after
imposing one of the conditions fulfilling Solar System tests,
(\ref{ablaether}) or (\ref{abkhronon}). Once ($\alpha$, $\beta$,
$\lambda$) are fixed, all coefficients in the field equations
(\ref{eqchixi}) and in the Einstein equations can be derived.

%%%%%%%%%%%%%%%%%%%%%%%
\subsection{Changing the gauge}
%%%%%%%%%%%%%%%%%%%%%%%

Reference \cite{Ma:1995ey} shows the impact of a gauge transformation
on the variables describing matter  and metric perturbations, and on
the form of their respective equations of evolution.  Here we show how
the fields $\chi$ and $\tilde{\xi}$ transform when switching from the
synchronous to the Newtonian gauge. This transformation is induced by
the particular change of coordinates
\be
x^0\mapsto x^0-\dot\beta(x,\tau)\;, 
\quad x^i\mapsto x^i-\pd_i \beta(x,\tau)-\epsilon^i(x)\;,
\ee
where
\be
\beta(x,\tau)=\int \di^3 k \frac{e^{i kx}}{2k^2}(h+6\eta)\;,
\quad
\d_i\epsilon^i=0\;.
\ee
After this transformation, the scalar part of the metric (\ref{FRW})
becomes diagonal,
\be
\label{Newmetric}
\di s^2=a(\tau)^2\left[-(1+2\psi)\di \tau^2+(1-2\phi)\di x^i \di x_i\right],
\ee
with
\be
\psi=\ddot\beta+\H\dot\beta\;,\quad\phi=\eta-\H\dot\beta\;.
\ee
The khronon and $\T$ field transform as
\be
\chi\mapsto\chi^N=\chi+\dot\beta\;,\quad
\tilde\xi\mapsto {\tilde\xi}^{N}=
\tilde\xi+\frac{\sqrt{8\pi G_{cos}}}{H_0}\dot{\Bar\Theta}\dot\beta\;.
\ee
Although the Boltzmann code {\sc class} features equations in both the
synchronous and Newtonian gauge, for simplicity, we choose to
implement the $\T$CDM equations in the synchronous gauge only.
However, when presenting the physical interpretation of numerical
results in the next section, we will refer to the evolution of
quantities in the Newtonian gauge, obtained by performing the above
transformation inside the code a posteriori (i.e. after solving the
equations of motion in the synchronous gauge).

%%%%%%%%%%%%%%%%%%%%%%%
\section{Observable effects}\label{sec:obser}
%%%%%%%%%%%%%%%%%%%%%%%

The modified evolution of linear perturbations in $\T$CDM leads to
observable consequences which we now discuss. We focus on the scalar
sector, known to contribute most to observable quantities.  The
effects of Lorentz violation on the tensor and vector sectors are
weakly constrained by current cosmological data
\cite{Nakashima:2011fu,ArmendarizPicon:2010rs,Lim:2004js}.  The only
effect on tensor modes is a small shift in their velocity
\cite{Jacobson:2008aj}.  The vector equations are identical in general
relativity and in the khronometric model, implying that vector
perturbations decay with time. Possible effects of vector modes in the
Einstein-aether model on CMB polarization have been discussed in
\cite{Nakashima:2011fu,ArmendarizPicon:2010rs}.   

At the qualitative level, one can identify three main effects
distinguishing the growth of perturbations in $\T$CDM from that in
$\Lambda$CDM. These are: 
{\it (i)} 
  a rescaling of the matter contribution in the Poisson equation 
(i.e., a different self-gravity of matter perturbations), 
{\it (ii)} 
  an additional contribution to the anisotropic stress, and 
{\it (iii)} 
  the presence of additional clustering species.
The first two effects are generic for any Lorentz violating
gravitational theory based on the Einstein-aether or khronometric
model\footnote{Similar effects are also known in other modified
  gravity and dark energy models, see e.g.
  \cite{Lue:2003ky,D'Amico:2012zv}.},
while the third is specific to the dynamical realization of dark
energy in $\T$CDM.

To understand these effects, let us work in the Newtonian gauge. The
Poisson equation (i.e., the sub-Hubble limit of the $(00)$ Einstein
equation) reads 
\be
\begin{split}
&k^2(2\phi-\alpha\, \psi)=-8\pi G_0 a^2 \sum_{\mathrm{other}} \rho_n \delta_n\\
&-\alpha k^2(\dot\chi+\HH (1-B)\chi)-\frac{\alpha\, G_0}{c^2_\Theta \tau_\a G_{cos}}
\left(H_\alpha \dot{\tilde \xi}-\frac{\psi}{\tau_a}\right)\;.
\label{eqNewsub}
\end{split}
\ee
Let us first concentrate on the contribution of standard matter. Using
the Friedmann equation (\ref{eq:Friedm}), the (time-dependent)
fraction of the total energy density of the Universe due to each
matter component is given by
\be
\label{fract}
f_n\equiv\frac{8\pi G_{cos}}{3H^2}\rho_n\;.
\ee
The Poisson equation takes the form
\be
\begin{split}
k^2 \phi &=
-4\pi G_N a^2 \sum_{\mathrm{other}} \rho_n \delta_n \\
&=
-\frac{3}{2}\frac{G_N}{G_{cos}}\HH^2 \sum_{\rm other}f_n\delta_n\;,
\label{eqNewsubma}
\end{split}
\ee
where for the time being we have omitted the terms in the second line
of (\ref{eqNewsub}) and neglected the anisotropic stress (i.e. we have
set $\psi=\phi$).  If we assume that the background evolution of the
Universe is standard, with 
${\cal H}$ and $f_n$ 
being exactly the same as in $\Lambda$CDM, Eq.~(\ref{eqNewsubma})
implies that the strength of the gravitational potential produced by
density perturbations is modified by the factor $G_N/G_{cos}$. When
this modified potential is substituted into the matter equations of
motion (which have the standard form), it leads to a different growth
rate of perturbations with respect to the $\Lambda$CDM case. For
instance, a straightforward calculation shows that during the matter
dominated epoch, the density contrast grows according to the modified
power-law (cf. \cite{Kobayashi:2010eh})
\be
\label{deltamat}
\delta\propto a^{\frac{1}{4}(-1+\sqrt{1+24 G_N/G_{cos}})}\;.
\ee
Notice that for small values of the parameters ($\a$, $\b$, $\l$) the
anomalous growth is proportional to
\be
\frac{G_N}{G_{cos}}-1= \frac{\Sigma}{2}+O(\alpha^2)\;,
\ee
where we have defined
\be
\label{sigma}
\Sigma\equiv\a+\b+3\l\;.
\ee
For $\Sigma=0$, the effect of modified self-gravity is strongly
suppressed. Hence we expect cosmological bounds on ($\a$, $\b$, $\l$)
to be weaker along this degeneracy direction. Incidentally, in the
Einstein-aether case $\Sigma$ is required to vanish (or, rather, be
extremely small) by the PPN constraints, see (\ref{ablaether}). This
explains why the cosmological bounds on Einstein-aether theory are
rather mild \cite{Zuntz:2008zz,Li:2007vz}. On the other hand, in the
khronometric model the PPN constraints are compatible with $\Sigma
\neq 0$ and the influence of Lorentz violation on cosmological
perturbations is more pronounced.

The second effect is understood from the tracefree part of the $(ij)$
Einstein equation,
\be
\begin{split}
k^2&(\psi-\phi)=\\
&-12\pi G_0 a^2\sum_{\mathrm{other}}( \rho_i +p_i)\sigma_i+
{\beta}k^2(\dot \chi+2\HH \chi).\label{eqanis}
\end{split}
\ee
Since the khronon field introduces a preferred direction $u_\mu$ in
space-time, it may generate anisotropic stress. This effect
(proportional to $\beta$) is accounted for by the last term in the
previous equation.  Generally speaking, adding anisotropic stress
amounts to increasing the viscosity of the cosmic fluid, and leads to
a damping of small-scale perturbations.

The third effect comes from gravitational interactions between
ordinary matter species and the scalar fields $\chi$ and
$\tilde{\xi}$. This interaction, described by the second line in
Eq.~(\ref{eqNewsub}), may play an important role under the condition
that the fields cluster and form sufficiently dense clumps. This might
be the case through a mechanism described in Ref.~\cite{Blas:2011en}.
The mixing between the dark energy perturbation $\tilde\xi$ and the
khronon $\chi$ (see Eqs.~(\ref{eqchixi})) gives rise to a mode whose
sound speed vanishes in the limit of small momentum $k$. This property
implies that the effective pressure associated with the mode is small,
and that the density perturbations ($\delta \rho_\chi$, $\delta
\rho_\xi$) can be amplified efficiently by gravitational collapse.
Hence, the $\T$CDM model features clustering dark energy. A
semi-quantitative analysis of this effect was performed in
\cite{Blas:2011en}, showing that structure formation is enhanced at
small comoving momenta (large wavelengths), $k\lesssim H_\a$.
Unfortunately, in this range, the quality of cosmological data is
rather poor, and this effect does not play a significant role in
actual observational constraints on the model. 

To illustrate how the above effects impact observable quantities, we
study numerically the evolution of cosmological perturbations in two
reference models using the Boltzmann code {\sc class}.  We will call
them the \emph{enhanced gravity} and the \emph{shear} models. The
corresponding parameter values are listed in Table~\ref{tab:1}.
Clearly, in the \emph{enhanced gravity} model we keep the effect {\it
(i)} while switching off the effect {\it (ii)}; in the case of the
\emph{shear} model the situation is opposite. The effect {\it (iii)}
is present in both models but we will see that it is always
subdominant. 
\begin{table}
\begin{center}
\begin{tabular}{|c|c|c|c|c|}
\hline
 Model & $\a$ & $\b$ & $\l$ &$\Sigma$\\\hline
\emph{enhanced gravity} & $0.2$ & $0$ & $0.1$&$0.5$\\
\hline
\emph{shear} & $0.05$ & $ 0.25$ & $-0.1$& 0\\\hline
\end{tabular}
\end{center}
\caption{Parameters for the \emph{enhanced gravity} and \emph{shear} models.}
\label{tab:1}
\end{table}
Note that the values in Table~\ref{tab:1} have been chosen very large
in order to make the modifications visible on the plots.  However,
these values are excluded by current data (see Sec.~\ref{sec:data}).

%%%%%%%%%%%%%%%%%%%%%%%%%%%%%%
\subsection{Effects on the CMB}
\label{ssec:obserA}
%%%%%%%%%%%%%%%%%%%%%%%%%%%%%%

In this subsection, we describe the changes induced by Lorentz
violation in the CMB temperature anisotropy spectrum.  We consider
$\T$CDM with the two reference sets of parameters listed above and
compare the results to $\Lambda$CDM. To highlight the changes, all
simulations are performed with  adjusted initial conditions such that,
in the limit $k\rightarrow 0$, the gravitational potential $\psi$ is
the same for all three models. For the standard cosmological
parameters, we choose the following values:
$n_s=1$, $h=0.7$, $\Omega_b=0.05$, $\Omega_{cdm}=0.25$, $A_s=2.3\times
10^{-9}$, $z_{reio}=10$.\\

\begin{figure}[h!]
  \begin{flushleft}
   \hspace{-1.3cm}
    \includegraphics[scale=0.35]{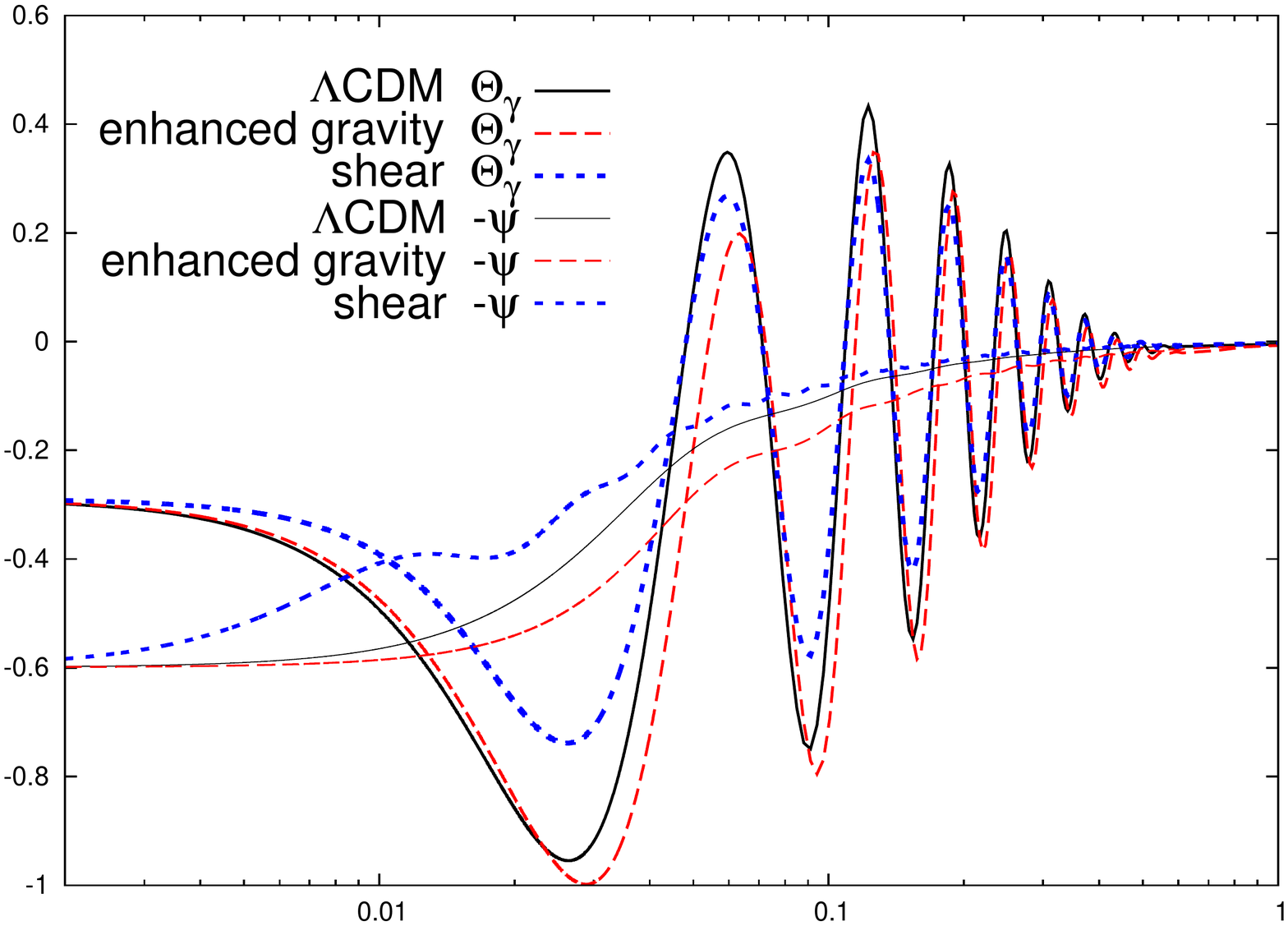}\\
    \vspace{-1.6cm}
    \hspace{-1.3cm}
    \includegraphics[scale=0.35]{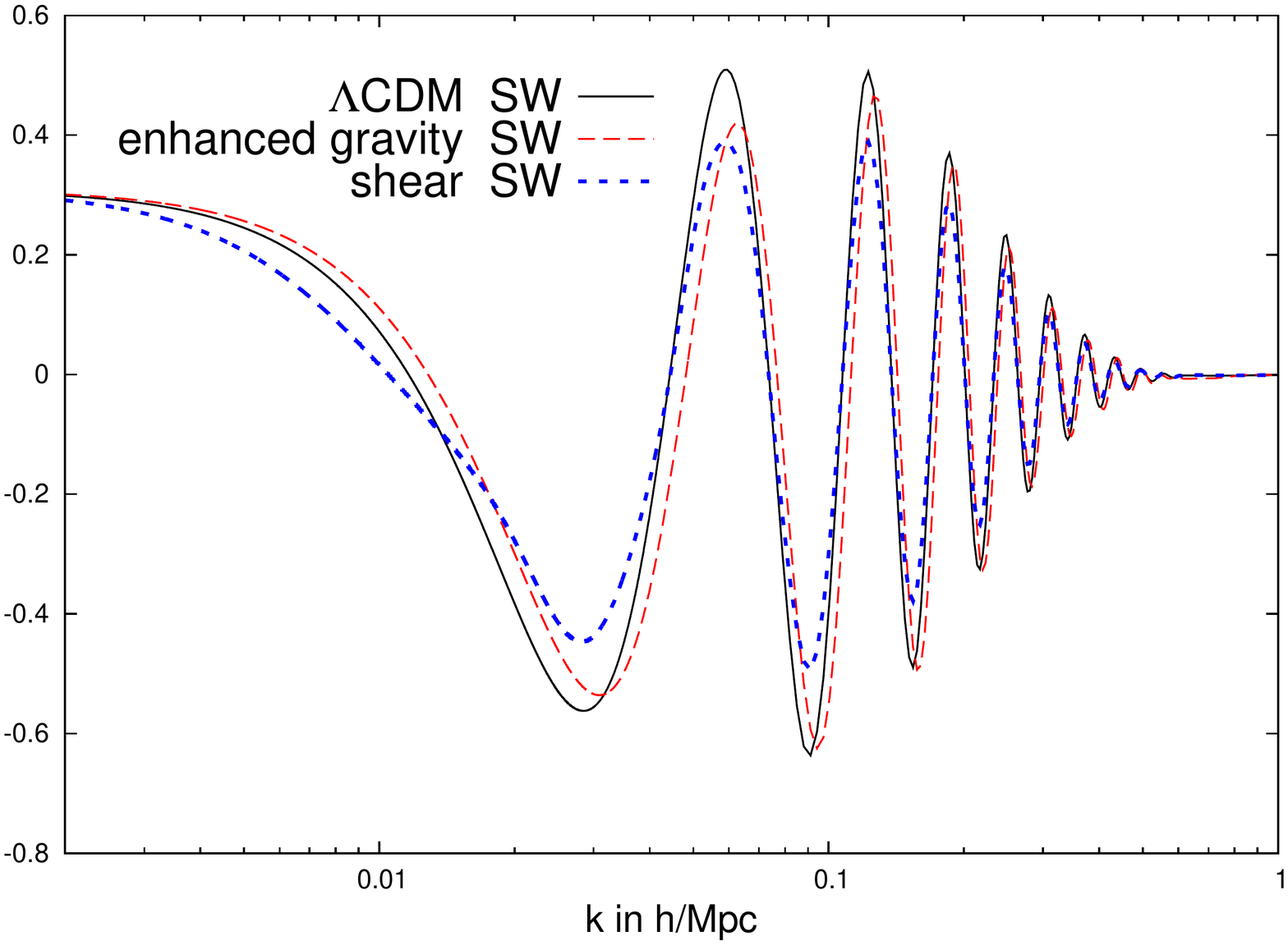}
  \end{flushleft}
  \vspace{-1.3cm}
  \caption{Top: photon temperature perturbation
    $\T_\gamma\equiv\delta_\gamma/4$ and gravitational potential
    $-\psi$ at decoupling, for the three models under scrutiny.
    Bottom: Sachs-Wolfe contribution, given by $\Theta_\gamma+\psi$
    (i.e. by the difference between the two curves above).}
  \label{fig:snapshots}
\end{figure}

\noindent
\emph{Enhanced gravity model}:
In Fig.~\ref{fig:snapshots} (top panel)
we can see that at the time of decoupling, the gravitational potential
in this model is enhanced (in absolute value) compared to
$\Lambda$CDM. 
This is due to an increase in the perturbation growth, governed by
$\Sigma$. On the same panel one observes two changes in the solution
for the photon temperature perturbations
$\T_\gamma\equiv\delta_\gamma/4$: a shift of the peaks of oscillations
towards higher momenta, and a shift of the zero point of oscillations.
These effects can be qualitatively understood from the combination of
the modified Poisson equation (\ref{eqNewsubma}) with the equations of
motion of the photon-baryon plasma before decoupling. The latter have
a standard form, and in the tight coupling approximation they reduce
to a single master equation,    
\begin{align}
 \ddot \Theta_{\gamma}+\frac{\dot R}{1+R}\dot \Theta_{\gamma}+k^2c_s^2\Theta_{\gamma} = -\frac{k^2}{3}\psi + \frac{\dot R}{1+R}\dot\phi+ 
 \ddot\phi,
  \label{eqn:ThetaPrimePrime}
\end{align}
where $R\equiv \frac{3\rho_b}{4\rho_\gamma}$ encodes the
baryon-to-photon density ratio, and $c_s\equiv (3(1+R))^{-1/2}$ is the
sound speed of density waves in the plasma in the absence of gravity.
According to Eq.~(\ref{eqNewsubma}), the first term on the r.h.s. of
(\ref{eqn:ThetaPrimePrime}) contains a contribution proportional to
$\T_\gamma$. This contribution is positive and larger than in
$\Lambda$CDM for $\Sigma>0$. It effectively decreases
the speed of sound in the plasma at the moment when each mode enters
inside the horizon\footnote{Strictly speaking, the Poisson equation
  (\ref{eqNewsubma}) is valid only for subhorizon modes. However, it
  is sufficient for our qualitative argument.},  
which translates into a shift of the peaks, observed in the top panel
of Fig.~\ref{fig:snapshots}.  Next, the zero point of the acoustic
oscillations is given by the value of $-\psi/3c_s^2$ at decoupling. As
already mentioned, $|\psi|$ is larger  in the {\it enhanced gravity}
model, leading to a shift in this zero point further away from
$\Theta_\gamma=0$. Finally, the amplitude of the acoustic
oscillations around the zero point is slightly smaller in the {\it
enhanced gravity} model than in $\Lambda$CDM.

The above features affect the Sachs--Wolfe (SW) contribution to
temperature anisotropies. In the bottom panel of
Fig.~\ref{fig:snapshots}, we plot this contribution, given by
$\Theta_{\gamma}+\psi$ at recombination.  One clearly sees the shift
of the peaks and notices that the even peaks are suppressed while the
odd ones are almost constant (with the notable exception of the first
peak, which is suppressed). This is due to the competition between
the different effects described in the previous paragraph.

\begin{figure}[h]
  \begin{flushleft}
    \hspace{-1.3cm}
    \includegraphics[scale=0.35]{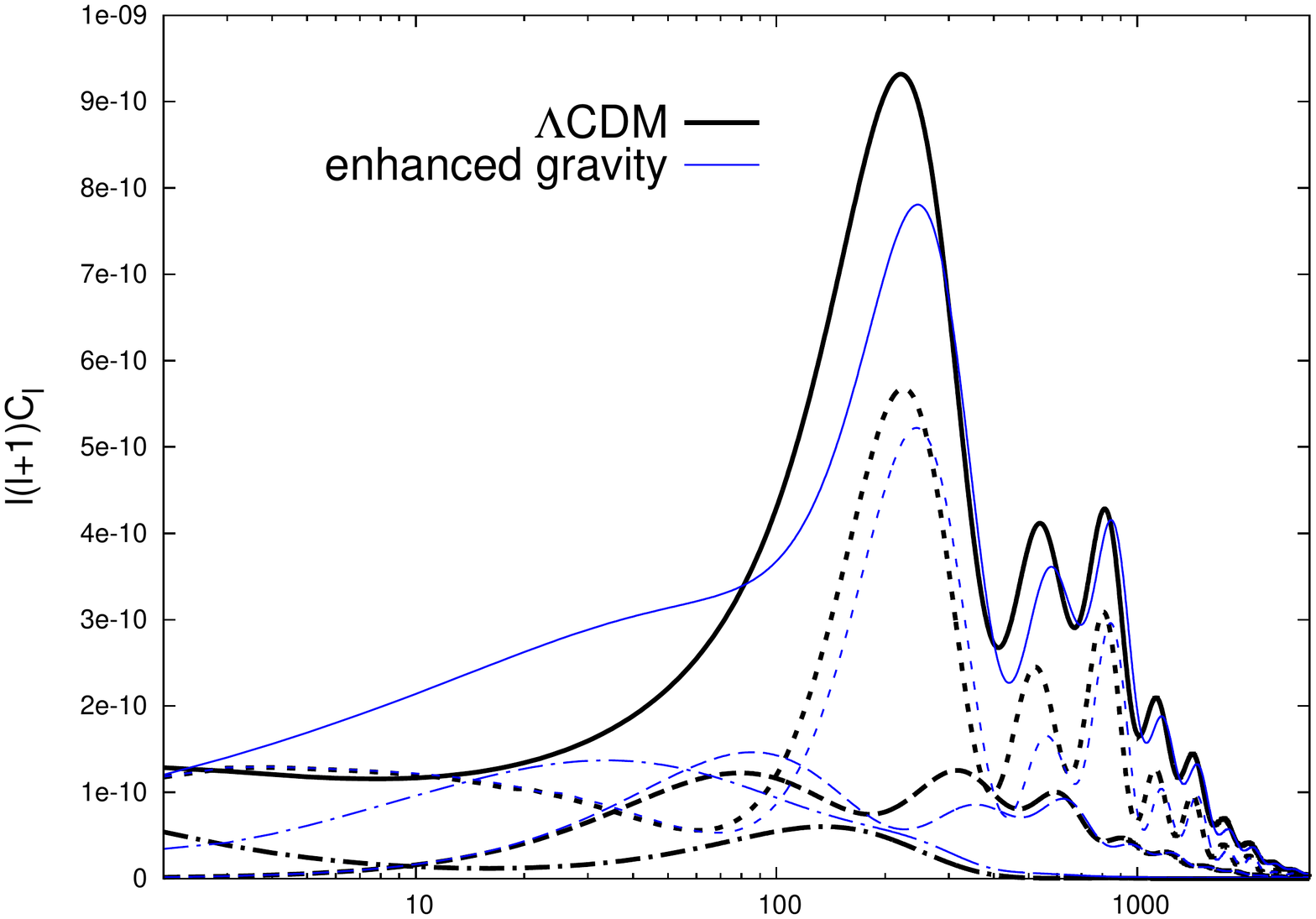}\\
    \vspace{-1.6cm}
    \hspace{-1.3cm}
    \includegraphics[scale=0.35]{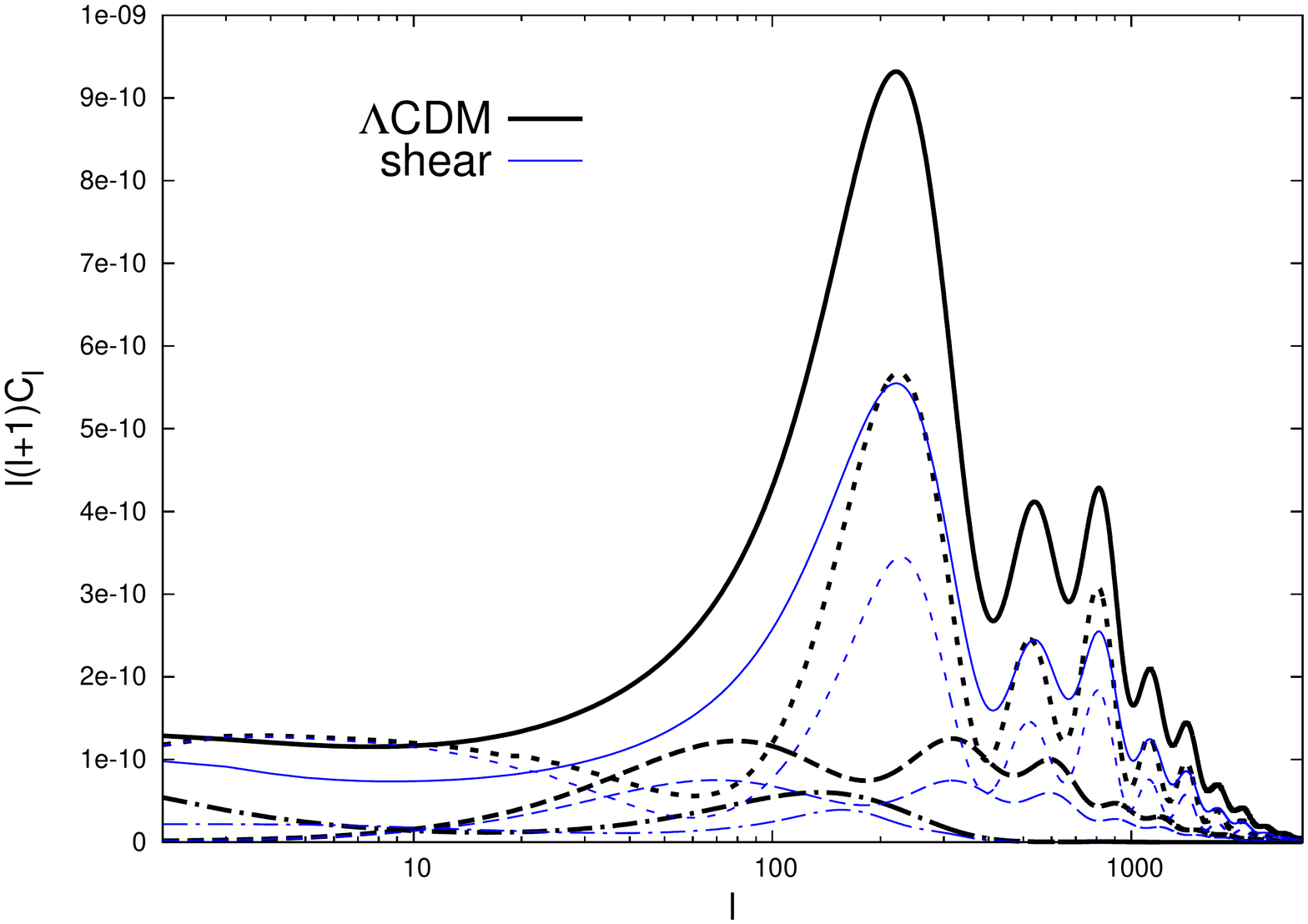}
  \end{flushleft}
  \vspace{-1.3cm}
  \caption{
     Temperature anisotropy spectrum (solid) and its decomposition in
     terms of Sachs--Wolfe (dotted), Doppler (dashed) and Integrated
     Sachs--Wolfe (dot-dashed) contributions. For clarity, we do not
     show cross correlations between these contributions.  Thick black
     lines represent the $\Lambda$CDM model, while thin blue lines are
     used for the two $\T$CDM reference models.}
  \label{fig:cl_decomposition}
\end{figure}

In the anisotropies observed today, the SW contribution is
supplemented by those coming from the Doppler and Integrated
Sachs--Wolfe (ISW) effects. The decomposition of the $C_\ell$ spectrum
in terms of these effects is presented in
Fig.~\ref{fig:cl_decomposition} for different models.  The comparison
of the \emph{enhanced gravity} model with $\Lambda$CDM is shown in the
top-panel.  In the SW contribution (as well as in the total spectrum),
we observe the expected suppression of the first, second and fourth
peaks, while the third peak amplitude is roughly unchanged.  On small
angular scales, the peaks are further suppressed by Silk damping.
Indeed, due to the shift in the phase of oscillations, they correspond
to smaller physical scales at recombination, that are more affected by
diffusion damping.  The Doppler effect, which depends on
$\dot\T_\gamma$ at recombination, is also modified. Finally, a
prominent feature clearly visible on the plot is the significant
enhancement of the ISW contribution in the range $10<l<100$, i.e.
between the regions usually affected by the early ISW effect
($100<l<200$) and the late ISW effect ($2<l<10$).  The ISW effect is
proportional to the time derivative of the gravitational potential. In
the $\Lambda$CDM model, the potential varies only during the epochs of
radiation and $\Lambda$ domination. However, in the \emph{enhanced
gravity} model, the growth of density perturbations entails a slow
increase of the gravitational potential also during the matter
dominated era, enhancing the ISW effect on a wide range of scales.

All in all, we conclude that the \emph{enhanced gravity} model
produces significant modifications in the spectrum of CMB
anisotropies. The pattern of these modifications is quite specific,
and apparently not degenerate with the effects of standard
cosmological parameters. \\

\noindent
\emph{Shear model}:
We recall that in this model, the $\chi$ field generates some
anisotropic stress and contributes to the shear of the perturbed
metric, as described by Eq.~(\ref{eqanis}).  The presence of shear
tends to smooth out metric perturbations on scales smaller than the
sound horizon associated with the sound speed of the $\chi$ field.
Note that for parameters of the particular \emph{shear} model studied
here, the field $\chi$ is superluminal\footnote{As pointed above, this
  does not present any inconsistencies in theories without Lorentz
  invariance.}
($c_\chi=\sqrt{3}$) and the suppression appears already on
super-Hubble scales.

This is indeed observed on the top panel of Fig.~\ref{fig:snapshots},
where the gravitational potential is clearly smaller (in absolute
value) compared to $\Lambda$CDM. It also exhibits notable wiggles
caused by oscillations in the $\chi$-field \cite{Blas:2011en}.  The
fact that $c_\chi$ is much larger than the photon-baryon sound speed
explains the shift between the phase of the oscillations seen in
$\psi$ and in $\Theta_\gamma$.  The suppression of $\psi$ shifts the
zero-point of the oscillations in the photon temperature $\T_\gamma$.
On the other hand, we do not see any shift in the positions of the
peaks, which is compatible with the previous discussion: the
self-gravity of radiation is not modified in this model.  
 
For fixed initial conditions $\Theta_\gamma(\tau_0)$, the amplitude of
the acoustic oscillations depends crucially on boosting effects,
imprinted around the time of Hubble crossing, and caused by the three
gravitational driving terms on the right-hand side of
Eq.~(\ref{eqn:ThetaPrimePrime}). In the \emph{shear} model, the
amplitude of acoustic oscillations is damped as a consequence of
smaller metric fluctuations and reduced gravitational boosting.  
This translates into an overall suppression
of the SW effect visible in the bottom panel of
Fig.~\ref{fig:snapshots}.

On Fig.~\ref{fig:cl_decomposition}, we see that the Doppler and ISW
contributions to the total temperature spectrum $C_\ell$ are also
lower in the \emph{shear} model than in $\Lambda$CDM. The net result
is a uniform suppression of all peaks.  One may expect that this
effect could be compensated, at least partially, by a rescaling of the
initial amplitude of perturbations. This suggests that pure
\emph{shear} models might be less constrained than \emph{enhanced
gravity} ones.

Our \emph{shear} model is similar to the one studied in
\cite{Zuntz:2008zz}, where it was claimed that the dominant effect on
the CMB comes through the ISW. Our analysis demonstrates that the
changes in the SW and Doppler contributions are equally important for
this model.

%%%%%%%%%%%%%%%%%%%%%%%%%%%%%%%%%%%
\subsection{Effects on the matter power spectrum}
%%%%%%%%%%%%%%%%%%%%%%%%%%%%%%%%%%%

Another observable affected by the behaviour of cosmological
perturbations in $\T$CDM is the matter power spectrum. We restrict
the discussion to low enough Fourier modes, $k\lesssim 0.1~$h/Mpc, for
which perturbations are still very close to the linear regime. The
study of non-linear clustering in this model is beyond the scope of
the present paper. 

The definition of what one calls the ``matter power spectrum'' is not
so obvious in $\T$CDM.  This model contains additional components that
contribute to perturbations of the total energy density.  One may
wonder whether they must be included in the calculation of the power
spectrum. In general, the answer to this question is yes. 
Indeed, the existing observations fall into two categories. The first
category probes directly metric perturbations (themselves related to
total density fluctuations): this is the case e.g. for cosmic shear
surveys or CMB lensing measurements. The second category measures the
clustering of compact objects like galaxies, halos or clusters. On
large scales, these objects are known to trace linearly the underlying
gravitational field~\cite{Bernardeau:2001qr}.  Thus, it appears
reasonable to \emph{define} the matter power spectrum using the
Poisson equation\footnote{Strictly speaking, this definition makes
  sense only for modes well inside the horizon, where the Poisson
  equation is valid.  However, this is sufficient for our study, because
  the power spectrum is actually measured for such modes only.}  
\be
\label{drhodef}
k^2\phi=-4\pi G_Na^2\delta\rho_{tot}\;.
\ee
Note that we have used here the locally determined value of the Newton
constant introduced in Eq.~(\ref{gn}). Comparing Eq.~(\ref{drhodef})
to Eq.~(\ref{eqNewsub}) and neglecting the anisotropic stress which is
very small on relevant scales, we see that the contribution of
ordinary matter to $\delta\rho_{tot}$ coincides with the standard
definition $\delta\rho_n$, while the contributions of the khronon and
$\tilde\xi$ fields read
\be\label{newtpois}
\begin{split}
&\delta\rho_\chi\equiv \frac{\alpha\, k^2}{8\pi a^2G_0}  (\dot \chi+\mathcal H (1-B)\chi)\;,\\
&\delta\rho_\xi\equiv  \frac{\a}{8\pi a^2
  G_{cos}c_\Theta^2\tau_\alpha}
\bigg(H_\a\dot{\tilde \xi}-\frac{\psi}{\tau_\a}\bigg)\;.
\end{split}
\ee  
We stress that all quantities here are taken in the Newtonian gauge. 

\begin{figure*}[!t]
  \begin{center}
    \includegraphics[scale=0.3]{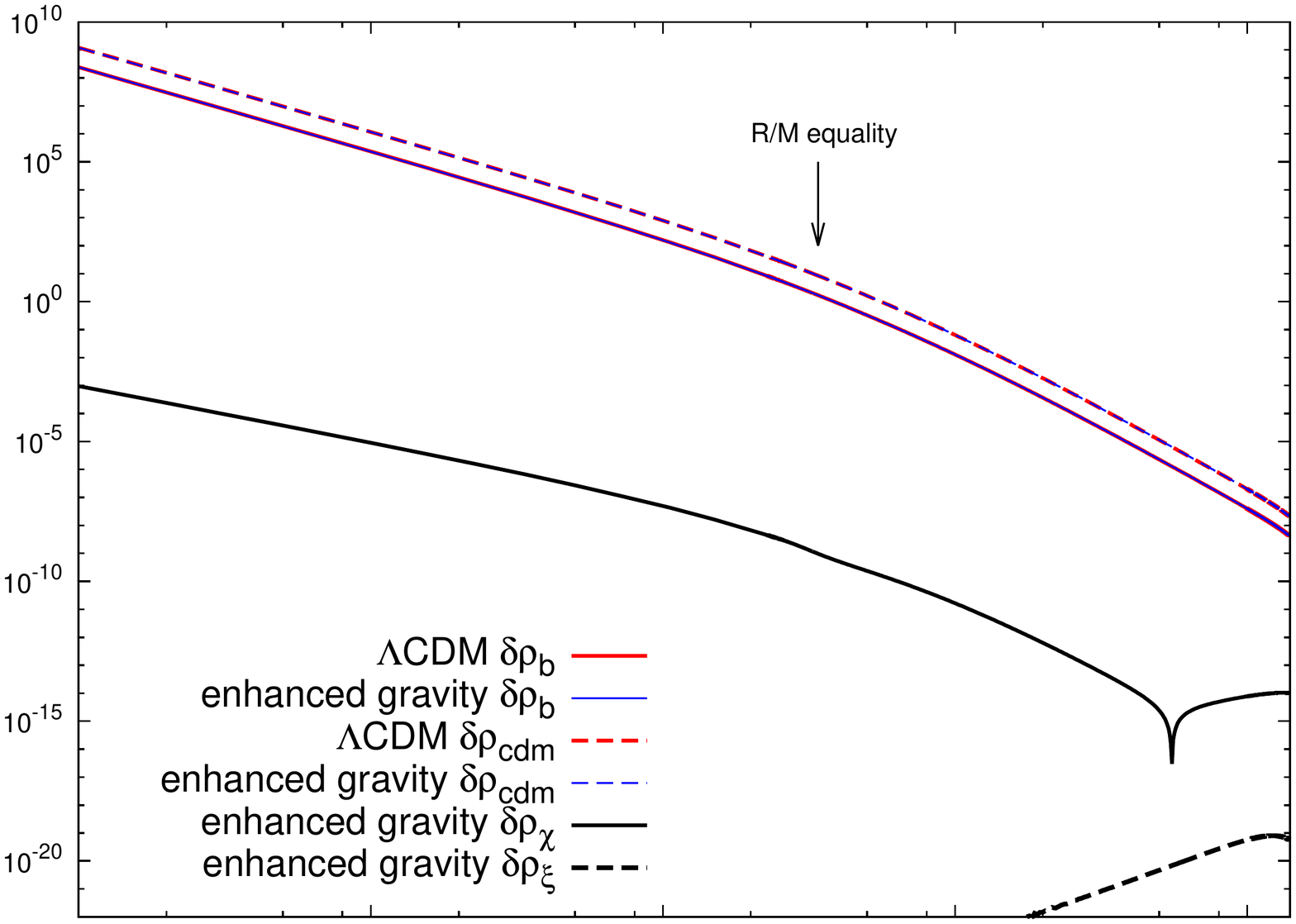}\hspace{-1cm}
    \includegraphics[scale=0.3]{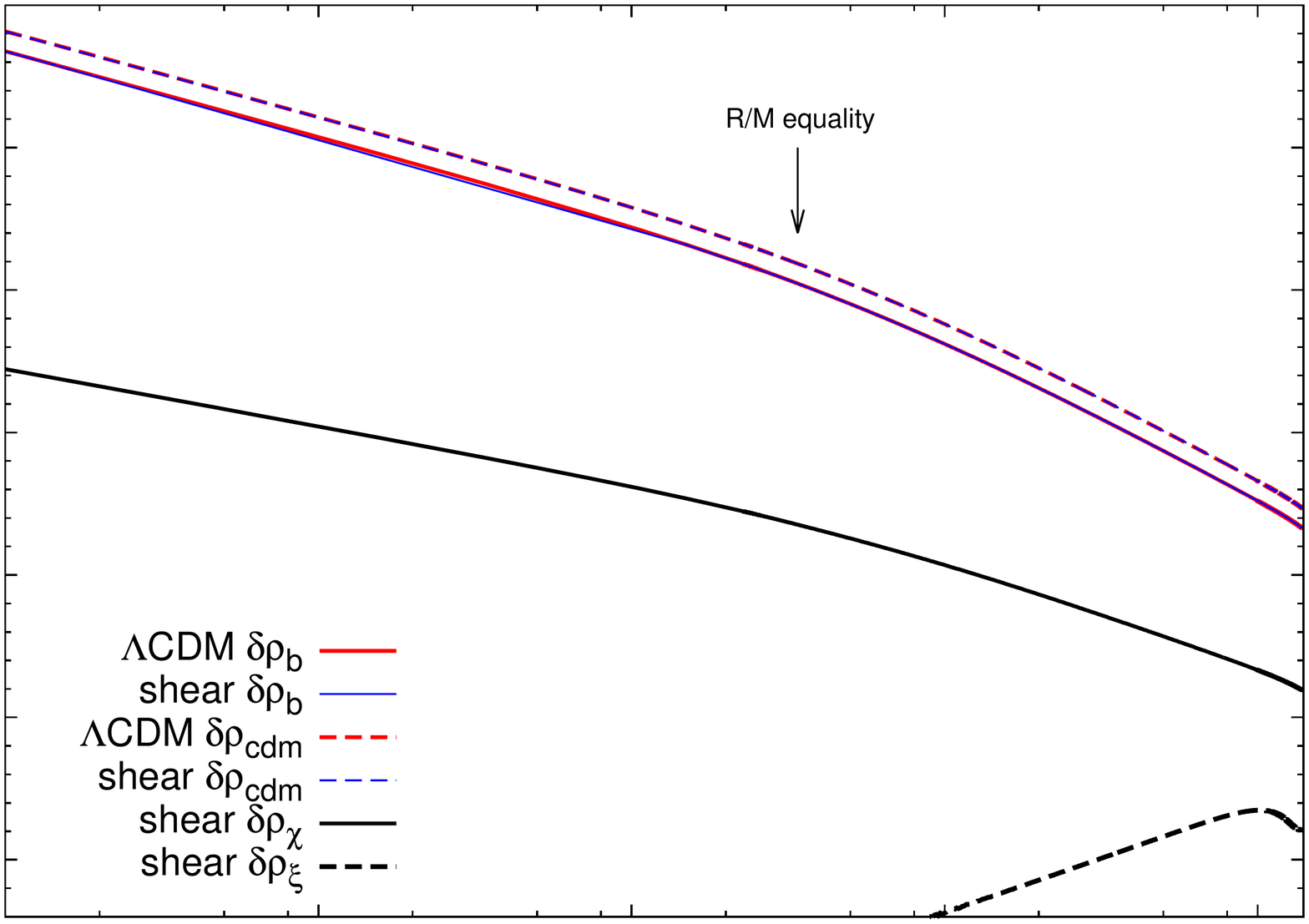}\vspace{-1cm}\\
    \includegraphics[scale=0.3]{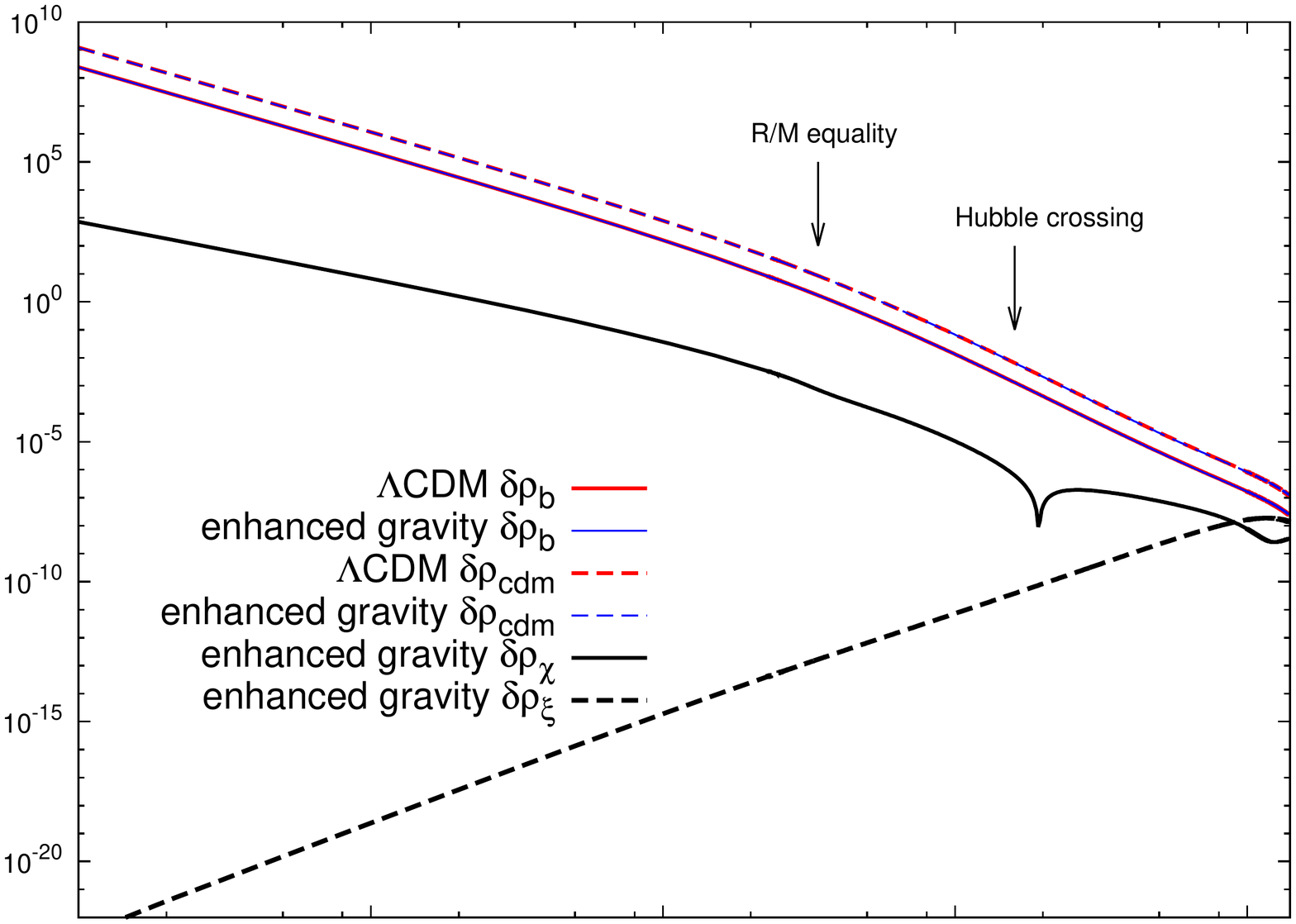}\hspace{-1cm}
    \includegraphics[scale=0.3]{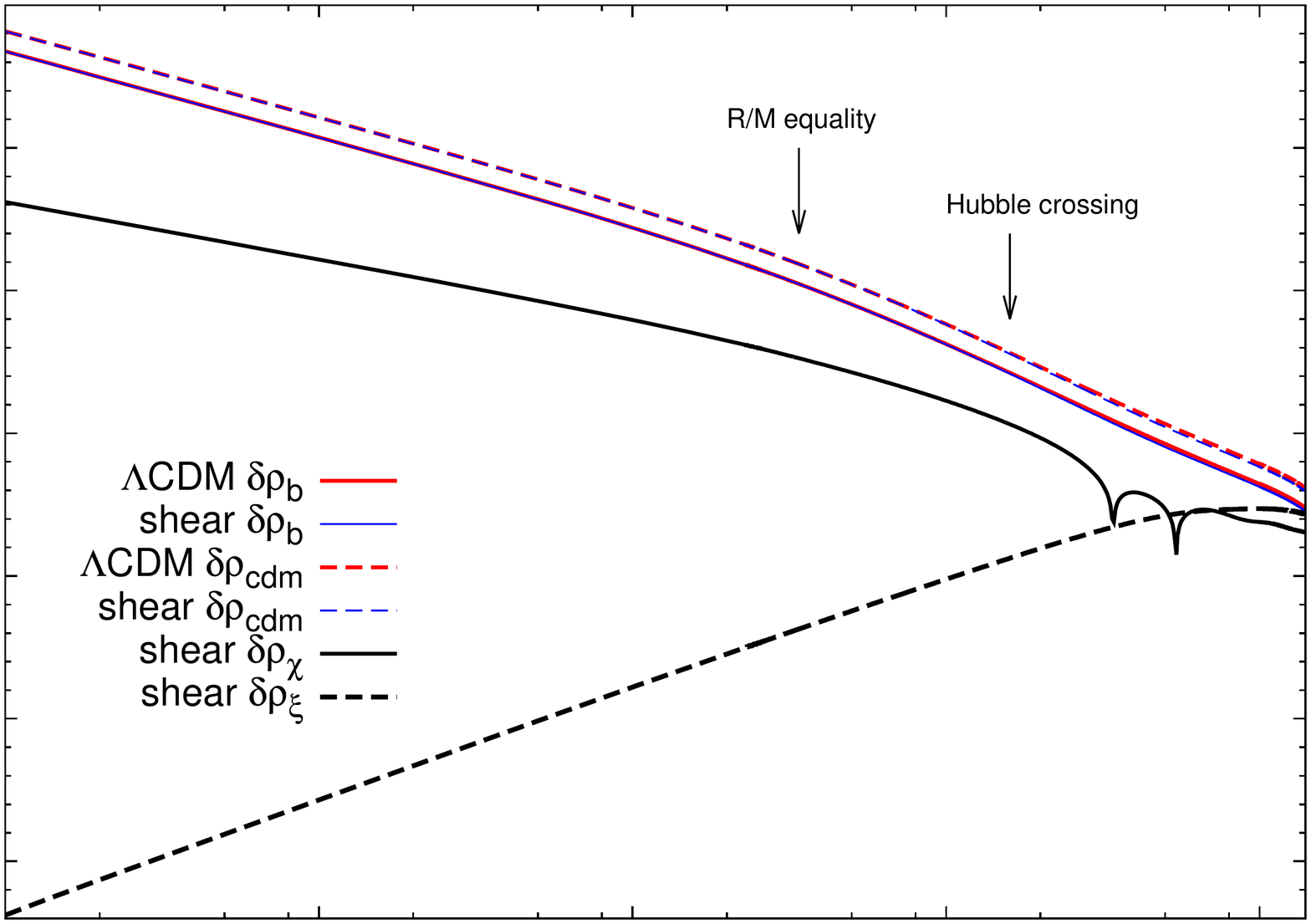}\vspace{-1cm}\\
    \includegraphics[scale=0.3]{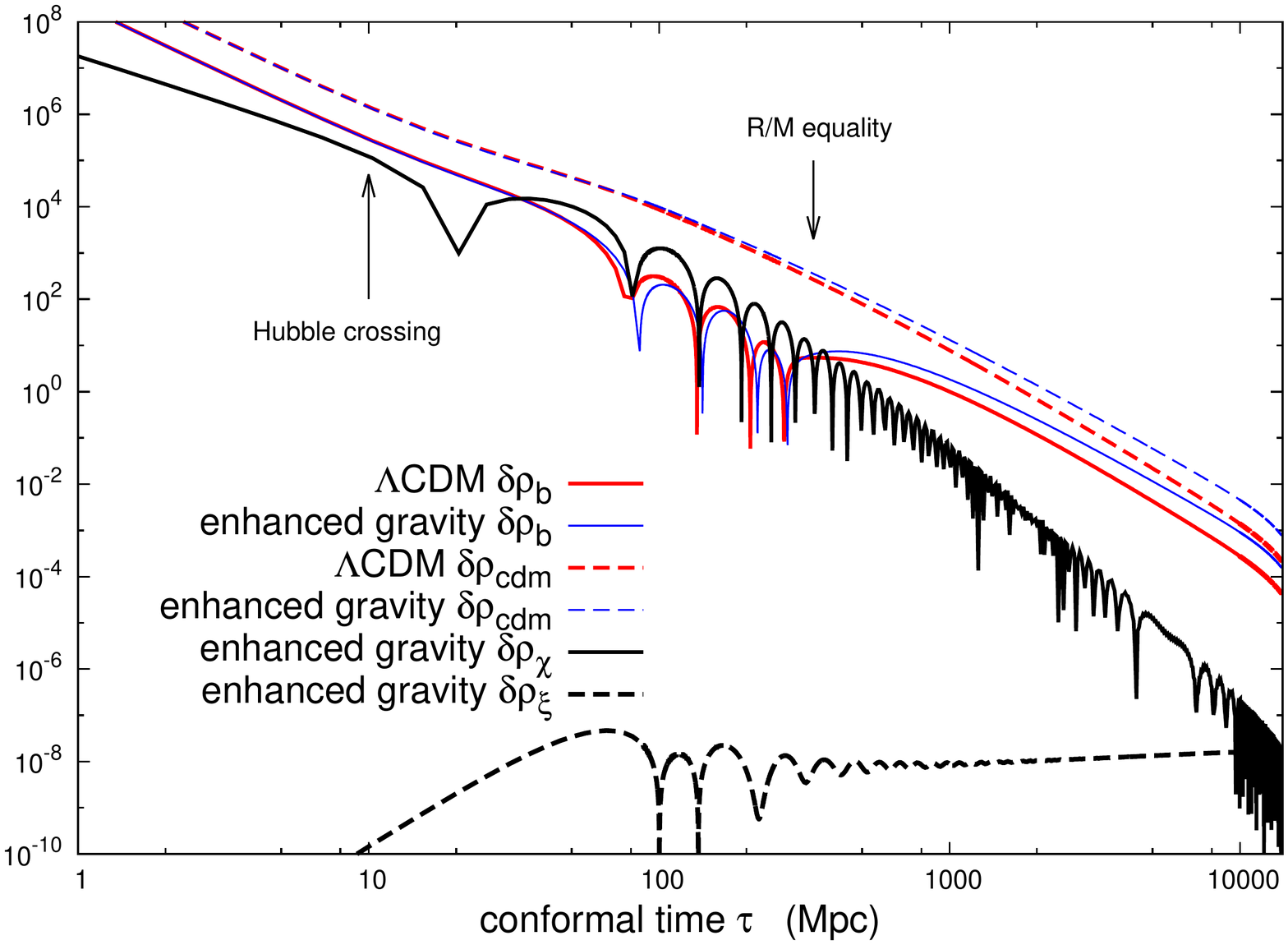}\hspace{-1cm}
    \includegraphics[scale=0.3]{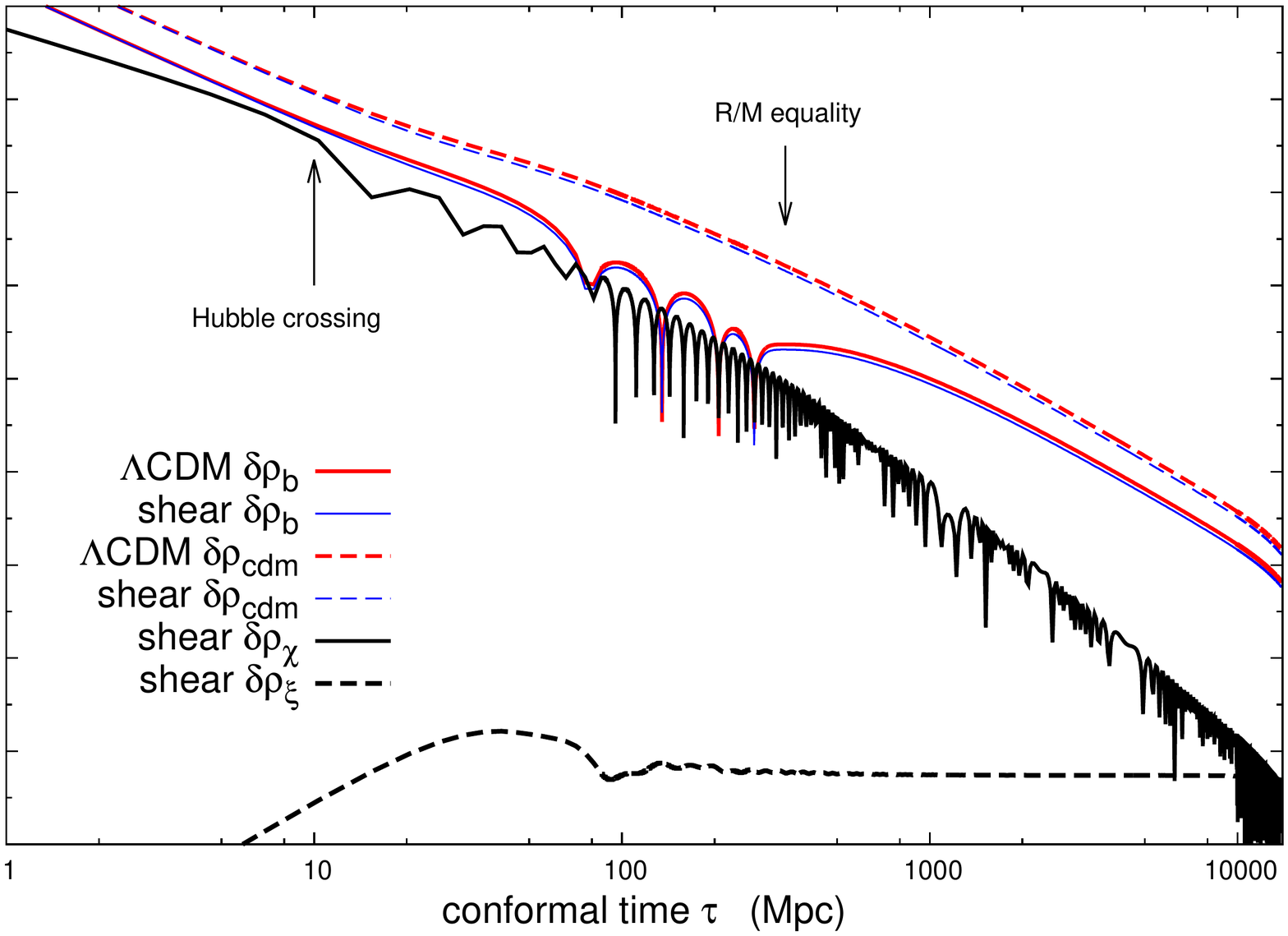}\vspace{-1cm}
  \end{center}
  \caption{ Evolution of the transfer function of various energy
    density perturbations $\delta \rho_i$ (arbitrary units) in the
    Newton gauge, for the \emph{enhanced gravity} model (left) and the
    \emph{shear} model (right), and for the three wavenumbers
    $k=10^{-7}$Mpc$^{-1}$ (top panels), $k=6\times 10^{-4}$Mpc$^{-1}$
    (middle panels) and $k=0.1$Mpc$^{-1}$ (bottom panels). For
    comparison, we show the evolution of $\delta \rho_b$ and $\delta
    \rho_{cdm}$ in the $\Lambda$CDM case. We also indicate the
    conformal time of Hubble crossing for the two largest 
wavenumbers,
    and the conformal time of radiation to matter equality.}
  \label{fig:delta_rhos}
\end{figure*}

To obtain the matter power spectrum, the density perturbation
$\delta\rho_{tot}$ must be divided by the total matter density (dark
matter plus baryons). The latter is determined from the background
cosmological solution. In the case of standard gravity, one would
again use the local Newton constant $G_N$ to infer the density from
the geometry.  Because in the $\T$CDM model the gravitational constant
in the Friedmann equation is different, the effective matter density
found in this way is renormalized compared to the actual value,
\be
\rho_\mathrm{eff}=\frac{G_{cos}}{G_N}(\rho_{cdm}+\rho_b)\;.
\ee
Thus we arrive at the following formula for the power spectrum,
\be
\label{Pkdef}
P(k)=
\bigg(\frac{G_N}{G_{cos}}\bigg)^2
\frac{\langle | \delta\rho_{tot}(\vec{k}) |^2 \rangle}{(\rho_{cdm}+\rho_b)^2}\;.
\ee
 Let us discuss the imprint of our two reference $\T$CDM models on this spectrum.
\\

\noindent
\emph{Enhanced gravity model}:
The left panels of Fig.~\ref{fig:delta_rhos} show the time dependence
of various contributions to $\delta\rho_{tot}$ in this model, for
three values of the momentum $k$. As expected, inside the Hubble
radius, matter density perturbations grow at a larger rate and are
enhanced compared to the $\Lambda$CDM case.  The contribution of the
$\chi$ and $\xi$ fields, though small, exhibits some interesting
features.  On super-Hubble scales, the $\xi$-density perturbation
rapidly grows with time.  This is due to the mixing between $\chi$ and
$\tilde{\xi}$ discussed in Ref.~\cite{Blas:2011en}, which gives rise
to a mode with vanishing sound speed at low momenta, and allows for
clustering of the dark energy.  In the case of modes crossing the
Hubble scale around the current epoch, $\delta\rho_\xi$ even becomes
comparable to the matter contributions at the present time.  However,
at shorter scales, the mixing between $\chi$ and $\tilde{\xi}$
disappears and the speeds of sound of these components become non
negligible\footnote{We remind that in this work we consider the regime
  where $c_\chi$, $c_\Theta$ are of order one. If instead these
  velocities were very small, the fields $\chi$ and $\tilde{\xi}$
could in principle cluster also on scales much smaller than the Hubble
radius.}. This leads to damped oscillations of these fields, clearly
visible in the lower left panel of Fig.~\ref{fig:delta_rhos}.
\begin{figure}[h]
  \begin{flushleft}
    \hspace{-1.25cm}
    \includegraphics[scale=0.35]{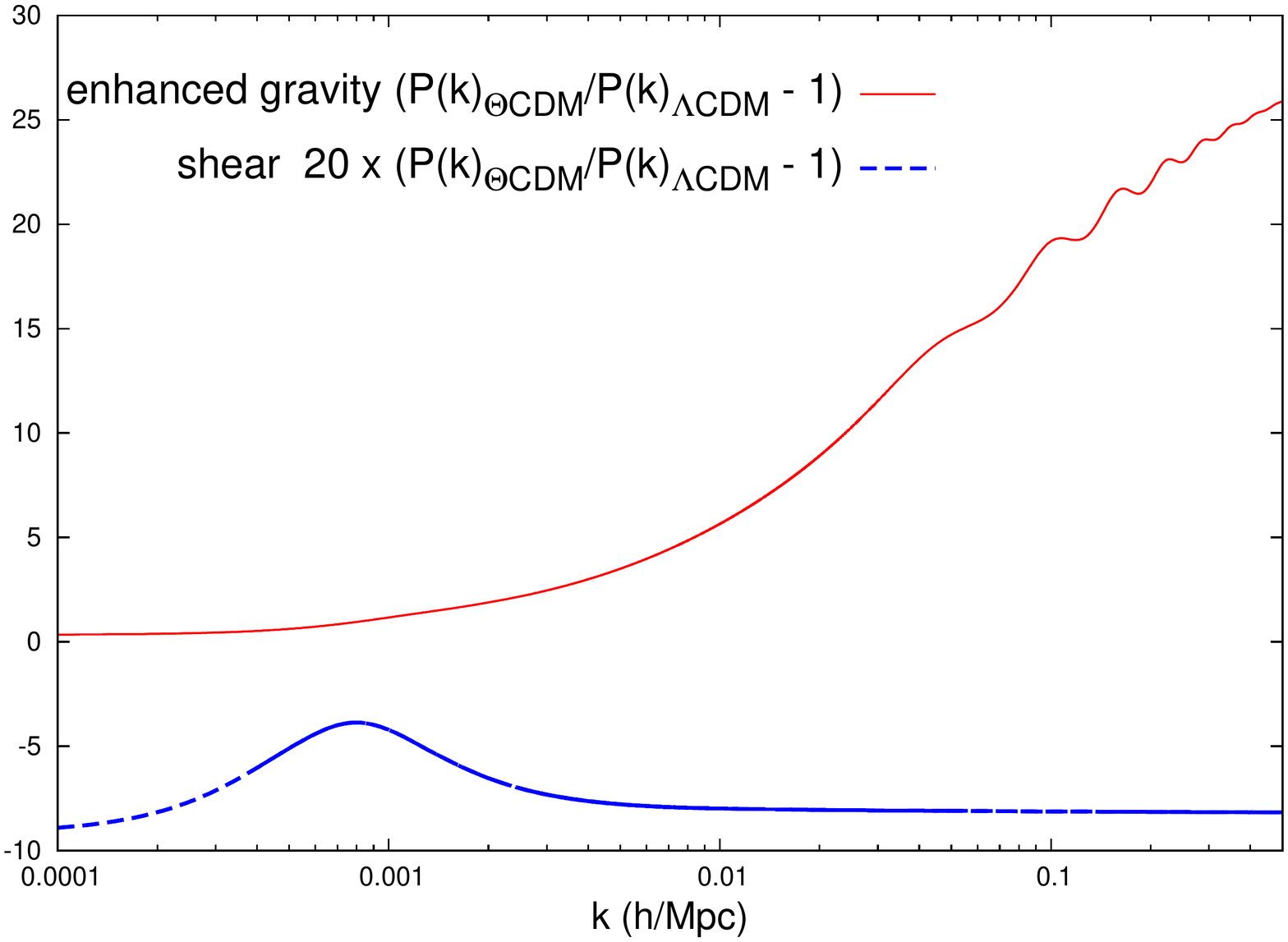}
    \vspace{-1.5cm}
  \end{flushleft}
  \caption{Ratios of the matter power spectra in the two reference
    $\T$CDM models and in $\Lambda$CDM at redshift $z=0$.}
  \label{fig:pk_ratio}
\end{figure}

The ratio of the power spectra in the \emph{enhanced gravity} model
and in $\Lambda$CDM is presented in Fig.~\ref{fig:pk_ratio}. The
accelerated growth of matter density perturbations translates into a
significant scale-dependent enhancement of the power spectrum on
scales that are below the Hubble radius today (i.e. with $k\gtrsim
0.0003\,h$/Mpc).  The curve exhibits small wiggles due to a shift in
the position of the peaks of baryon acoustic oscillations (cf. the
shift of CMB peaks discussed in Sec.~\ref{ssec:obserA}).  On
sub-Hubble scales, the contribution of $\delta\rho_\chi$ and
$\delta\rho_\xi$ to the total matter power spectrum is
negligible\footnote{As mentioned above, this conclusion could change
in a scenario where a tiny value of $c_\chi$ or $c_\Theta$  would be
assumed.}.  Finally, let us point out that the curve in
Fig.~\ref{fig:pk_ratio} must be taken with a grain of salt at
$k\gtrsim 0.1\,h$/Mpc, where non-linearities become important.\\ 

\noindent
\emph{Shear model}: 
Our definition of the matter power spectrum refers to sub-Hubble
wavelengths only, and also uses the assumption that $\phi$ and $\psi$
are equal in Eq.~(\ref{eqNewsub}), which leads to Eq.~(\ref{drhodef}).
Hence we need to check whether the anisotropic stress (responsible for
a possible difference between the two metric perturbations, see
Eq.~(\ref{eqanis})) vanishes inside the Hubble radius in the shear
model. It is well-known that the anisotropic stress of neutrinos and
of decoupled photons decays inside the Hubble radius, due to
free-streaming. This conclusion also applies to the anisotropic stress
of the khronon, due to the dynamics of the $\chi$ field. For instance,
during matter domination, $\chi$ oscillates with an envelope decaying
like $\tau^{-3}$ on sub-Hubble scale, implying that its averaged
energy density decays like $\delta \rho_\chi \propto \tau^{-8} \propto
a^{-4}$. The contribution of the khronon anisotropic stress to the
difference $(\phi-\psi)$ is then oscillating with an envelope
proportional to $\tau^{-4} \propto a^{-2}$. At the same time, $\phi$
and $\psi$ are almost constant. Hence the relative difference between
$\phi$ and $\psi$ quickly becomes negligible, and our definition of
the matter power spectrum on sub-Hubble scale is applicable.

If we normalize perturbations to the same initial
value of $\psi$, the evolution of ($\delta\rho_{cdm}$, $\delta\rho_b$)
is identical above the Hubble scale in the $\Lambda$CDM and
\emph{shear} models. However, in the \emph{shear} case, the
anisotropic stress of the khronon cannot be neglected soon after the
time of Hubble crossing.  This leads to a smoothing of metric
perturbations and to a suppression of $\delta\rho_{cdm}$ and
$\delta\rho_b$ that is clearly visible in the lower right panel of
Fig.~\ref{fig:delta_rhos}. Well after Hubble crossing, the khronon
anisotropic stress becomes negligible and ($\delta\rho_{cdm}$,
$\delta\rho_b$) evolve like in $\Lambda$CDM, but with a constant
offset coming from the suppression experienced soon after Hubble
crossing.

This suppression is seen better in Fig.~\ref{fig:pk_ratio}. For
the chosen parameter values, the effect of the \emph{shear} model  is
smaller than for the \emph{enhanced gravity} model. In order to see
clearly both effects with the same scale, we multiplied the difference
between the power spectra in the \emph{shear} model and in
$\Lambda$CDM by 20. Inside the Hubble radius, the suppression of the
power spectrum in the \emph{shear} model is almost scale-independent.
When approaching the current value of the Hubble scale, the
suppression is reduced by the counteracting effect of additional
density fluctuations in the $\xi$ field. However, this reduction is
only significant on scales that are too large to be observed with
good precision (due to the sampling variance associated to a given
survey). Besides, on such scales our definition of the matter power
spectrum is no longer 
applicable\footnote{Like in the   \emph{enhanced
gravity} model, we note that in a scenario with tiny values of
$c_\Theta$ or $c_\chi$, density fluctuations in the $\tilde{\xi}$ or
$\chi$ field could in principle be significant on much smaller
scales.}.

We reach the same conclusion as in the CMB case: on observable scales,
the effect of pure \emph{shear} models can be mimicked by an overall
reduction of the primordial fluctuation amplitude, which suggests that
these models are more weakly constrained than \emph{enhanced gravity}
models.

%%%%%%%%%%%%%%%%%%%%%%%%%%%%%%%%%%%%%%%%
\section{Comparison with current data}\label{sec:data}
%%%%%%%%%%%%%%%%%%%%%%%%%%%%%%%%%%%%%%%%

We will now compare the khronometric $\Theta$CDM model to CMB  and LSS
data, using the parameter inference code {\sc Monte
Python}\footnote{\tt http://montepython.net} \cite{Audren:2012wb}.
For the CMB, we use here WMAP 7-year data \cite{Komatsu:2010fb}, and
SPT data from 2008 and 2009 \cite{Keisler:2011aw}.  Updating our
analysis with recent Planck data \cite{Ade:2013ktc} could give a small
improvement on parameter constraints, without changing their order of
magnitude. For LSS, we rely on galaxy power spectrum data from the
WiggleZ redshift survey \cite{Parkinson:2012vd}. We choose to perform
our runs in the synchronous gauge. 

We include in the fit eight free cosmological parameters, which
are the usual six free parameters  of the minimal flat $\Lambda$CDM
model, plus $\beta$ and $\beta+\lambda$ (this combination is chosen to
facilitate the convergence of the chains).  The parameter $\alpha$ is
fixed to $2\beta$, which is the condition for satisfying all bounds
coming from Solar System tests in the khronometric model\footnote{We
  decided to focus only on the khronometric case, since from the
  comments after Eq.~\eqref{sigma} it is clear that the bounds are
  stronger in this model than in the Einstein-aether case. This is
  confirmed by comparing our results with the constraints presented in
  Ref.~\cite{Zuntz:2008zz}.}.
As far as cosmology is concerned, this condition is actually not
necessary, but we implement it anyway in order to deal only with
realistic models.  We also vary three nuisance parameters describing
the foreground contamination of SPT data, and we marginalize over
these parameters following strictly the approach of
Ref.~\cite{Keisler:2011aw}.

\begin{figure*}
  \includegraphics[scale=0.45]{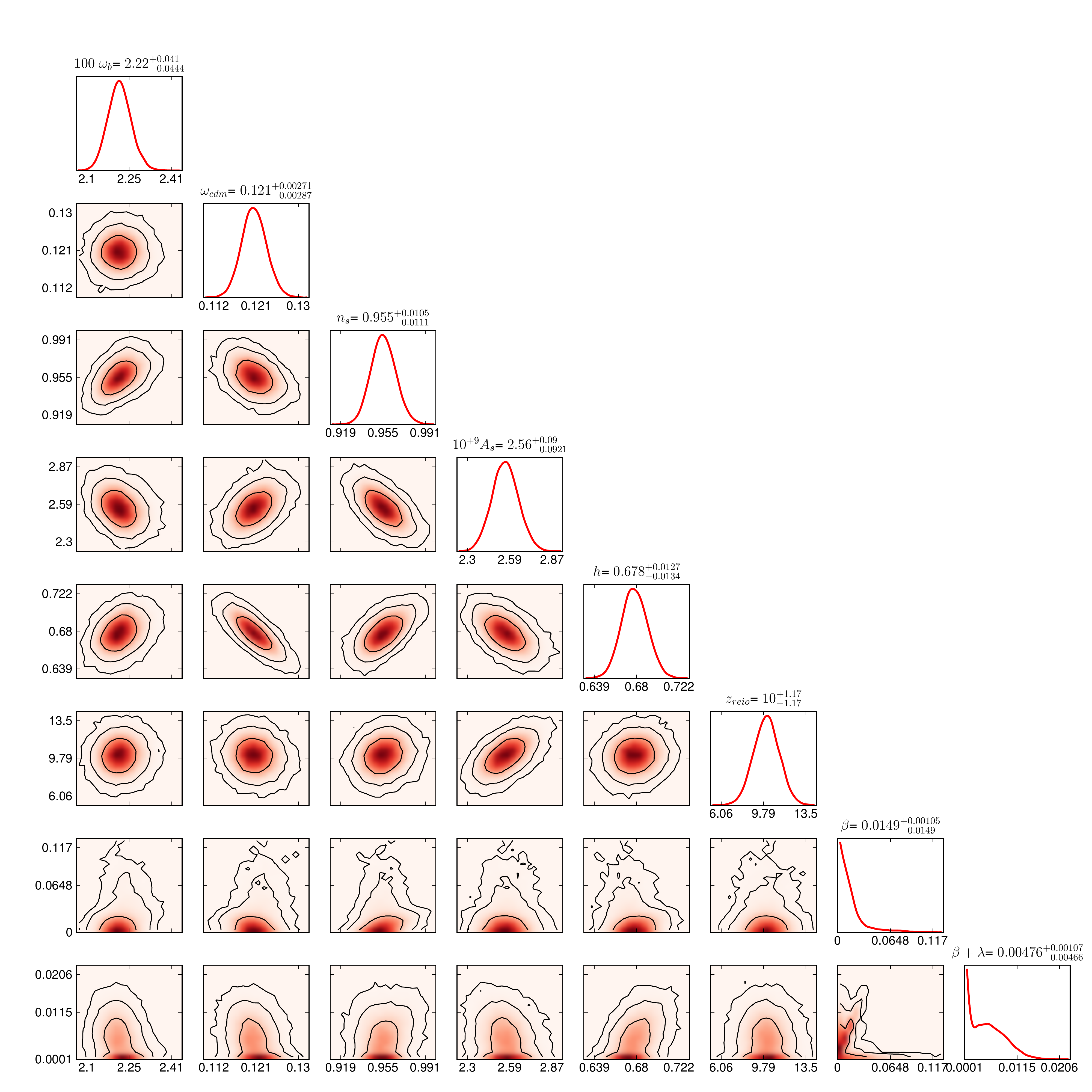}
  \caption{Triangle plot showing the one-dimensional marginalized
    posterior distribution and the two-dimensional probability contours
    (at the 68, 95 and 99\% confidence level) of the parameters of the
    khronometric $\Theta$CDM model.}
\label{fig:triangle}
\end{figure*}

We show our results for the Bayesian minimum credible interval of
each parameter in Table~\ref{table}. The one-dimensional and
two-dimensional posterior parameter distributions are displayed in
Fig.~\ref{fig:triangle} (omitting the three SPT nuisance parameters
for clarity).

\begin{table*}
  \begin{tabular}{|cccccccc|}
     \hline$100~\omega_{b }$  & $\omega_{cdm }$  & $n_{s }$  &
     $10^{+9}A_{s }$  & $h$  & $z_{reio }$  & $\beta$  &
     $\beta+\lambda$  \\ \hline 
     \vspace{-0.1cm}
     &&&&&&&\\
     $2.219_{-0.044}^{+0.041}$ & $0.1205_{-0.0029}^{+0.0027}$ &
     $0.9552_{-0.011}^{+0.011}$ & $2.556_{-0.092}^{+0.09}$ &
     $0.6782_{-0.013}^{+0.013}$ & $10.05_{-1.2}^{+1.2}$ &
     $<0.05$ & $<0.012$ \\[0.2cm]
     \hline
   \end{tabular}
\caption{Mean values and 68\% confidence limits of the minimum credible
interval of the parameters of the khronometric $\Theta$CDM model. For $\beta$ and
$\beta+\lambda$, we show instead 95\% upper limits.}
\label{table}
\end{table*}

The $\Theta$CDM parameters  are found to be weakly correlated with the
six standard model parameters, and strongly correlated with  each
other. Isoprobability contours in the $(\beta, \beta+\lambda)$ space
have a complicated shape. They are very elongated along two directions
of degeneracy, corresponding to $\beta+\lambda \simeq 0$ and $\beta
\simeq 3/2(\beta+\lambda)$. Away from these two special directions,
the contours are more regular, and closer to an ellipse centered on
the origin. The posterior probability peaks at $(\beta,
\beta+\lambda)\simeq (0,0)$, showing that the data brings no evidence
in favor of the $\Theta$CDM model.  The two directions of degeneracy
can be interpreted as follows. 

First, the case $\beta+\lambda \simeq 0$ corresponds to  $\Sigma
\simeq 0$ for our choice of $\alpha$. This is the region of pure
\emph{shear} models, discussed in the previous subsection. We already
argued that the differences with $\Lambda$CDM are less important in
this case\footnote{Under the condition $\alpha=2\beta$, the velocity of the
  $\chi$-field is given by $c^2_\chi=\Sigma/(6\beta)$. Thus, the limit
  $\Sigma=0$ formally corresponds to vanishing sound speed of the
  $\chi$-component. One could expect that this would lead to an amplification of
  the $\chi$-perturbations and, as a consequence, to strong constraints on the
  model. However, in practice, $c_\chi$ never becomes smaller than $\simeq
  10^{-2}$ during the Monte Carlo sampling of the parameters and the clustering
  of $\chi$ does not appear to be significant.}
than in the \emph{enhanced gravity} case corresponding to
the orthogonal direction $\beta=0$. This explains the factor four
difference between constraints on $\beta$ and on $\beta+\lambda$ in
Table~\ref{table}.

The other case has a less straightforward interpretation. The relation
$\beta \simeq 3/2(\beta+\lambda)$ implies that the khronon has a
squared sound speed $c_\chi^2 \simeq 1/3$, similar to that of
neutrinos and of tightly coupled photons. Hence, for these models, the
khronon does not introduce a new characteristic scale, and the 
oscillations seen in the lower left panel of Fig.~\ref{fig:delta_rhos} may 
 mimic those of photons or neutrinos, thus lowering the
sensitivity along this direction. We checked explicitly with the
modified {\sc class} code that in this particular case, the sum of all
effects described in the previous section (enhanced gravity,
additional shear and extra clustering species) can be nearly cancelled
at the level of CMB anisotropies by a variation of the standard
cosmological parameters. This situation is analogous to the case of
extra relativistic degrees of freedom, for which the impact of a small
variation of $N_\mathrm{eff}$  on the CMB can be partially compensated
by a variation of other parameters, leading to a well-known degeneracy
between $N_\mathrm{eff}$ and $H_0$~\cite{book}. $\Theta$CDM models
with $c_\chi^2 \simeq 1/3$ are not exactly equivalent to models with
extra relativistic degrees of freedom, but their effect can be
compensated in a similar way. This explains the weak correlation
observed in Fig.~\ref{fig:triangle} between standard parameters and
$\Theta$CDM parameters. By comparing with a run based on CMB data
only, we find that the inclusion of LSS data helps in breaking this
degeneracy (by limiting enhanced gravity effects on the matter power
spectrum), but even in presence of WiggleZ data, the degeneracy
appears very clearly in Fig.~\ref{fig:triangle}.

%%%%%%%%%%%%%%%%%%%%%%%
\section{Conclusions}\label{sec:concl}
%%%%%%%%%%%%%%%%%%%%%%%

In this work, we described the impact of the $\T$CDM model  on
cosmological observables, on scales where linear cosmological
perturbation theory is valid.  This model is an alternative to the
$\Lambda$CDM scenario in which the acceleration of the universe
expansion is caused by a dynamical scalar field $\Theta$.  An
important difference between this model and most quintessence models
is that the action of the field $\Theta$ is naturally protected from
large ultraviolet corrections.  In addition, the $\T$CDM model is
embedded in a family of gravitation theories with broken Lorentz
invariance, which may have an ultraviolet completion in the framework
of Ho\v rava gravity.  The latter theories are characterized by a
time-like dynamical vector field $u^\m$ which defines a preferred time
direction. Typical examples are the Einstein-aether and khronometric
theories. The  $\T$CDM model is the simplest extension of these
theories to include a mechanism for cosmic acceleration.  Under simple
constraints on its parameters, it passes all local tests of gravity:
Solar System dynamics, gravity wave emission, black hole structure.
We have shown that the $\Theta$CDM model may produce observable
effects on cosmological scales, constrained by current cosmological
data, but still potentially detectable with future ones.

The background evolution in $\T$CDM is identical to $\Lambda$CDM.  The
only potential difference could come from a component playing the role
of stiff matter, but in order to preserve primordial nucleosynthesis,
we must assume that this component is too small for playing any role
in late-time cosmology, and for affecting CMB and large scale
structure observables. However, 
the  evolution of cosmological perturbations is
 generically
very different in the $\T$CDM and $\Lambda$CDM models.  We studied the
evolution equations of scalar perturbations in $\T$CDM and identified
three different effects: 
{\it (i)} 
a renormalization of the matter contribution to the Poisson equation 
(i.e. a different self-gravity of matter perturbations), 
{\it (ii)} 
a new contribution to the anisotropic stress, and 
{\it (iii)} 
the presence of additional clustering degrees of freedom. The first
two effects are generic for Lorentz violating gravitation theories
based on a unit time-like vector, while the third one is specific to
the dynamical realization of dark energy in $\T$CDM. 

We implemented the equations of $\T$CDM in the Boltzmann code  {\sc
class} \cite{Blas:2011rf}, in order to compute accurately the impact
of the three effects {\it (i)} , {\it (ii)} , {\it (iii)}. We found
that they affect the power spectrum of Cosmic Microwave Background
anisotropies in a very particular way (shift in the position and
amplitude of the peaks, and enhanced ISW), as can be seen in
Fig.~\ref{fig:cl_decomposition}. Furthermore, they influence the
shape of the matter power spectrum (different amplitude and slope on
observable scales, shift in the position of baryon acoustic
oscillations).  These effects are shown in
Figures~\ref{fig:delta_rhos} and \ref{fig:pk_ratio}.

To derive constraints on the free parameters of the khronometric
$\T$CDM model (considering only parameter combinations satisfying
Solar System tests), we used data from WMAP (7-year), from SPT (2008
and 2009 observations), and from the WiggleZ redshift survey.  We ran
the parameter inference code {\sc Monte Python} \cite{Audren:2012wb}
and found the bounds displayed in Table~\ref{table} and illustrated in
Fig.~\ref{fig:triangle}.  Quite remarkably, these bounds constrain
deviations from general relativity to better than the percent level.
They are stronger than those from several tests of gravitation, such
as radiation damping of binary systems, or BBN.  They are also
stronger for the khronometric case studied in this paper than for the
Einstein-aether model considered in Ref.~\cite{Zuntz:2008zz}, due to
the absence of the effect {\it (i)} in the latter case.  The vanishing
of {\it (i)} in the Einstein-aether theory (once Solar System bounds
are imposed) seems to be simply a coincidence.

The fact that our bounds are dominated by the effect {\it (i)} -- modified
self-gravity of the matter perturbations -- suggests that they may have a wider
application. The bound on the combination $\beta+\lambda$ in Table II can be
cast into the constraint on the discrepancy between the gravitational constants
appearing in the Newton law and in the Friedmann equation, $|G_N/G_{cos}-1| <
0.018$ at $95 \%$ confidence level.  Though in our model we cannot completely
disentangle {\it (i)} from other effects, we believe that the above constraint
will apply, at least by order of magnitude, to any theory predicting a
time-independent discrepancy between $G_N$ and $G_{cos}$. It
represents an improvement compared to the bounds existing in the literature
\cite{Robbers:2007ca,Bean:2010zq}.  

Our analysis can be extended in several ways.  First, we have only
used the scalar sector of the theory. For the Einstein-aether case,
the theory contains also propagating vector modes, whose influence on
B-type CMB polarization could provide further observational
tests~\cite{ArmendarizPicon:2010rs,Nakashima:2011fu}.  Second, the
analysis can be updated with CMB data from Planck, and in the future,
with large scale structure data from DES, LSST or Euclid.  It would
also be extremely interesting to go beyond the linear regime and try
to understand the consequences of $\T$CDM for non-linear structure
formation.  Finally, to completely characterise possible deviations
from Lorentz invariance in the context of cosmology, one can consider
the option that this symmetry is violated also in the dark matter
sector~\cite{Blas:2012vn}.

\section*{Acknowlegements}
We are grateful to Mikhail Ivanov, Gregory Gabadadze, Roman
Scoccimarro and Takahiro Tanaka for useful discussions. D.B. and S.S.
thank the organizers and participants of the Kavli IPMU Focus Week on
Gravity and Lorentz Violations for the encouraging interest and
valuable comments.  S.S. thanks the Center for Cosmology and Particle
Physics of NYU for hospitality during the completion of this work.
This work was supported in part by the Grant of the President of
Russian Federation NS-5590.2012.2 (S.S.), the Russian Ministry of
Science and Education under the contract 8412 (S.S.), the RFBR grants
11-02-01528 (S.S.), 12-02-01203 (S.S.) and by the Dynasty Foundation
(S.S.). J.L. and B.A. acknowledge support from the Swiss National Science
Foundation.

\appendix
%%%%%%%%%%%%%%%%%%%%%%%%%%%
\section{Initial conditions}\label{sec:app}
%%%%%%%%%%%%%%%%%%%%%%%%%%%%

The initial conditions for the system of differential equations
describing the evolution of a given Fourier mode $k$ are set in the
radiation era, at a moment $\tau_0$  when the wavelength associated to
$k$ is well outside the Hubble scale, i.e. when $k\ll \H$. As
discussed in  \cite{Kobayashi:2010eh,Blas:2011en}, under broad
assumptions, the dominant contribution to the perturbations is
provided by the growing adiabatic mode. In this Appendix we determine
this mode analytically on super-Hubble scales using the equations of
Sec.~\ref{ssec:cosper}. 

Taking into account that $a\propto \tau$ during radiation domination
($\tau$ stands for conformal time), and keeping only leading order
terms in an expansion in $(k/\H) \approx (k\tau)$, we can simplify
Eqs.~(\ref{eq:chi}) and (\ref{eq:xi}):
\begin{align}
\label{eq:chisH}
&\ddot\chi+\frac{2}{\tau}\dot\chi+\frac{2B}{\tau^2}\chi
-\frac{G_0}{\Gc}\frac{H_\a}{\tau_\a}\tilde\xi
+\frac{c_\chi^2}{2}\dot h+\frac{2\b}{\a}\dot\eta=0\;,\\
\label{eq:xisH}
&\ddot{\tilde\xi}+\frac{2}{\tau}\dot{\tilde\xi}=0\;.
\end{align}
On the other hand, Eqs.~(\ref{eq:T00}), (\ref{eq:Tii}) can be rewritten as
\begin{align}
\label{eq:etasH}
\eta=&\frac{G_0}{2k^2\Gc}\frac{\dot h}{\tau}
-\frac{3}{2k^2\tau^2}\frac{G_0}{\Gc}\big(R_\nu\delta_\nu+(1-R_\nu)\delta_\gamma\big)\nonumber\\
&-\frac{1}{2}\bigg[\a\dot\chi+\a(1-B)\frac{\chi}{\tau}\bigg]
-\frac{\a}{2k^2c^2_\T}\frac{G_0}{\Gc}\frac{H_\a}{\tau_\a}\dot{\tilde\xi}\;,\\
\label{eq:hsH}
\ddot h+&\frac{\dot h}{\tau}
+\frac{6}{\tau^2}\big(R_\nu\delta_\nu+(1-R_\nu)\delta_\gamma\big)=\nonumber\\
&-\frac{\Gc}{G_0}k^2\a(1+B)\bigg(\dot\chi+\frac{\chi}{\tau}\bigg)
-\frac{\a H_\a\dot{\tilde\xi}}{c_\T^2\tau_\a}\;,
\end{align}
where $R_\nu$ is the ratio of the density of neutrinos to the total
radiation density.  Since we have not modified the matter sector of
the theory,  the solutions for matter perturbations in the
super-Hubble regime coincide with Eqs.~(92) of \cite{Ma:1995ey}.  Note
that we have left aside the remaining pair of Einstein equations
(\ref{eq:etadot}), (\ref{eq:tracefree}), that are redundant and
automatically satisfied by the solution
(\ref{eq:ansatzsH})---(\ref{eq:nonstandsi}).

At initial time $\tau_0$, the $\T$-field contribution to the total
energy density is negligible.  It is natural to expect that its
fluctuations can also be neglected, which corresponds to dropping off
the terms containing $\tilde \xi$ in Eqs.~(\ref{eq:chisH}),
(\ref{eq:etasH}) and (\ref{eq:hsH}). Let us find the precise
conditions under which  this is possible.  The non-decaying solution
of (\ref{eq:xisH}) is 
\be
\label{eq:xisi}
\tilde\xi=\tilde\xi_0\;,
\ee 
where $\tilde\xi_0$ is a constant implying that the $\tilde \xi$
contribution drops out of Eqs.~(\ref{eq:etasH}), (\ref{eq:hsH}). As
for Eq.~(\ref{eq:chisH}), we find that the $\tilde\xi$-term is
negligible provided the following condition is satisfied:
\be
\label{conds1}
\tilde\xi_0\ll\frac{\tau_\a \tau_0}{H_\a}h_0\;,
\ee
where we have used the expression (\ref{ansatzsHa}) for $h$ and have
assumed $c_\Theta,c_\chi\sim 1$.  This inequality is the condition for
having negligible isocurvature perturbations in the $\tilde\xi$-field.
It is likely to be satisfied under plausible assumptions about the
origin of the primordial $\tilde\xi$-perturbations \cite{Blas:2011en}.
However, this statement must be taken with a grain of salt: a detailed
theory of the $\T$-field dynamics starting from the inflationary epoch
is required to put it on the solid ground. We leave developing such a
theory for the future. In the present work we will set $\tilde\xi_0$
to zero in the numerical simulations.

For other fields we use the ansatz\footnote{In fact, $h_1, \
  \delta_{\gamma1}$ and $\delta_{\nu1}$ are not necessary to specify
  the initial conditions for the numerical procedure. However, they
  must be taken into account to obtain the expression for the
  coefficient $\eta_1$.}
(cf. \cite{Ma:1995ey}):
\bseq
\label{eq:ansatzsH}
\begin{align}
\label{ansatzsHa}
&h=h_0 \tau^2+h_1 \tau^4\;,\quad\eta=\eta_0+\eta_1 \tau^2\;,\\
&\delta_\gamma=\delta_{\gamma\,0} \tau^2+\delta_{\gamma\,1} \tau^4\;,
\quad\delta_\nu=\delta_{\nu\,0} \tau^2+\delta_{\nu\,1} \tau^4\;,\\
&\theta_\gamma=\theta_{\gamma\,0} \tau^3\;,
\quad\theta_\nu=\theta_{\nu\,0} \tau^3\;,\\
&\sigma_\nu=\sigma_{\nu\,0} \tau^2\;,
\quad\chi=\chi_0 \tau^3\;.
\end{align}
\eseq
From the matter equations (Eqs.~(92) of \cite{Ma:1995ey}) we obtain
the standard relations:
\bseq
\label{eq:standsi}
\begin{align}
&\delta_{\gamma\,0}=-\frac{2}{3}h_0\;, 
\quad\delta_{\gamma\,1}=\frac{k^2}{54}h_0-\frac{2}{3}h_1\;,\\
&\delta_{\nu\,0}=-\frac{2}{3}h_0\;, 
\quad\delta_{\nu\,1}=\frac{k^2}{30}h_0+\frac{4k^2}{45}\eta_1
-\frac{2}{3}h_1\;,\\
&\theta_{\gamma\,0}=-\frac{k^2}{18}h_0\;,
\quad\theta_{\nu\,0}=-\frac{k^2}{10}h_0-\frac{4k^2}{15}\eta_1\;,\\
&\sigma_{\nu\,0}=\frac{2}{15}h_0+\frac{4}{5}\eta_1\;.\label{eq:icss}
\end{align}
\eseq
To find the relations between the fields $h_0,h_1,\eta_1$ we use
(\ref{eq:etasH}), (\ref{eq:hsH}) and (\ref{eq:chisH}).  In the leading
order the first equations is satisfied identically, while the second
yields
\be
\label{eq:eta0si}
\eta_0=\frac{2G_0}{k^2\Gc}h_0\;.
\ee
Considering the subleading order in (\ref{eq:etasH}), (\ref{eq:hsH})
and the leading order in (\ref{eq:chisH}) we obtain
\bseq
\label{eq:nonstandsi}
\begin{align}
&\chi_0=-\frac{c_\chi^2}{2(6+B)}h_0-\frac{2\b}{\a(6+B)}\eta_1\;,\\
&h_1=\bigg[-\frac{k^2(5+4R_\nu)}{540}+\frac{\a\,
  k^2c_\chi^2(1+B)}{6(6+B)}
\frac{\Gc}{G_0}\bigg]h_0
\nonumber\\
&\quad\quad\quad-\bigg[\frac{2k^2R_\nu}{45}
-\frac{2k^2\b(1+B)}{3(6+B)}\frac{\Gc}{G_0}\bigg]\eta_1\;,\\
&\eta_1=-\frac{h_0}{6}\frac{5+4R_\nu
-\frac{45}{2}\a\, c_\chi^2\frac{\Gc}{G_0}}{
15\frac{\Gc}{G_0}(1-\b)+4R_\nu}\;.
\end{align}
\eseq
It is now straightforward to check that the obtained solution
satisfies the remaining Einstein's equation (\ref{eq:etadot}),
(\ref{eq:tracefree}).

To summarize, the adiabatic mode is given by Eqs.~(\ref{eq:ansatzsH})
with the coefficients related by (\ref{eq:standsi}),
(\ref{eq:eta0si}), (\ref{eq:nonstandsi}) to the single constant $h_0$
setting the overall normalization.  To these expressions we add the
standard adiabatic initial conditions for baryons and dark matter
\cite{Ma:1995ey}:
\be
\delta_{cdm}=\delta_b=\frac{3}{4}\delta_\gamma\;,~~~\theta_{cdm}=0\;,
~~~\theta_b=\theta_\gamma\;.
\label{eq:adia}
\ee

We cross-checked the validity of our initial conditions using the
numerical code. When fixing initial conditions at $\tau_0$, we find
that over an extended range of time above $\tau_0$, the numerical
solutions  for $(h, \eta, \chi)$ and for all density perturbations
remain equal to their analytic expressions in
Eqs.~(\ref{eq:ansatzsH})-(\ref{eq:nonstandsi}). Hence, the above
initial conditions correctly describe the attractor solution of the
full system of equations that is compatible with the adiabatic
condition (\ref{eq:adia}) for matter fields. At leading order, the
attractor solution for the field  $\tilde{\xi}$ is given by
\begin{align}
\tilde{\xi} = \frac{(c_\Theta k \tau)^2}{42 H_\alpha \tau_\alpha} \chi
\end{align}
%
%\noindent 
and evolves proportionally to $\tau^6$. Since this solution is too
small to back-react on other fields, it makes no difference to fix
the initial $\tilde{\xi}$ to the above solution or to zero (in the
latter case,  $\tilde{\xi}$ reaches very quickly the attractor
solution).

When deriving initial conditions, we assumed full radiation
domination, and neglected all corrections related to the contribution
of non-relativistic matter to the expansion rate. Hence our initial
conditions are accurate only if imposed at very early time. We
decreased $\tau_0$ in the code until getting very stable results. This
is the case for the value $\tau_0 =10^{-2}$Mpc, that we adopted in all
simulations.

%%%%%%%%%%%%%%%%%%%%%%%%%%%%%%%%%%%

\end{document}